\documentclass[12pt]{article}

\usepackage{amsmath,amssymb,epsfig,subfigure,color,soul,fleqn,rotating}
\usepackage{ulem,slashed}
\usepackage{cite}
\usepackage[colorlinks,citecolor=blue,urlcolor=blue,linkcolor=blue]{hyperref}
\usepackage{xcolor}

\newcommand{\be}{\begin{eqnarray}}
\newcommand{\ee}{\end{eqnarray}}
\newcommand{\nee}{\nonumber\end{eqnarray}}
\newcommand{\nn}{\nonumber\\}

\newcommand{\mch}[1] {m_{\ti \x^+_{#1}}}
\newcommand{\mnt}[1] {m_{\ti \x^0_{#1}}}

\newcommand{\msg}    {m_{\ti g}}
\newcommand{\msu}[1] {m_{\ti u_{#1}}}
\newcommand{\msd}[1] {m_{\ti d_{#1}}}

\def\gev             {{\rm GeV}}

\def\be            {\begin{equation}}
\def\ee            {\end{equation}}
\def\bea            {\begin{eqnarray}}
\def\eea            {\end{eqnarray}}

\definecolor{mybrown}{cmyk}{0,0.9,1.5,0.3}

\def\a              {\alpha}
\def\b               {\beta}
\def\d               {\delta}

\def\x               {\chi}
\def\ti              {\tilde}
\def\sq              {\ti q}
\def\st              {\ti t}
\def\sb              {\ti b}
\def\ch              {\ti \x^\pm}

\def\nt              {\ti \x^0}
\def\sg              {\ti g}

\def\su                {\ti{u}}

\def \sca                 {\ti{c}}
\def\sd                {\ti{d}}
\def\ss                  {\ti{s}}

\def\dll            {\d^{LL}_{23}}
\def\durr            {\d^{uRR}_{23}}
\def\durl            {\d^{uRL}_{23}}
\def\dulr            {\d^{uLR}_{23}}
\def\ddrr            {\d^{dRR}_{23}}
\def\ddrl            {\d^{dRL}_{23}}
\def\ddlr            {\d^{dLR}_{23}}

\newcommand{\AddrGAKUGEI}{%
 \it Department of Physics, Tokyo Gakugei University, Koganei,
Tokyo 184-8501, Japan\\}

\newcommand{\AddrHEPHY}{%
 \it Institut f\"ur Hochenergiephysik der \"Osterreichischen Akademie
der Wissenschaften, A-1050 Vienna, Austria\\}

\title{Two-body decays of gluino at full one-loop level \\in the quark-flavour violating MSSM}

\author{H. Eberl${}^1$, E. Ginina${}^{1}\footnote{On leave of absence from the Institute of Nuclear Research and Nuclear Energy, Sofia. }$, K.~Hidaka${}^{2}$}
\date{
\small $^1$ \AddrHEPHY 
          $^2$ \AddrGAKUGEI  
}

\definecolor{darkgreen}{rgb}{0,.5,0}

\begin{document}

\begin{flushright}
HEPHY-PUB 984/17\\
\end{flushright}
\begingroup
\let\newpage\relax
\maketitle
\endgroup

\maketitle
\thispagestyle{empty}

\begin{abstract}

We study the two-body decays of the gluino at full one-loop level in the Minimal Supersymmetric Standard Model with quark-flavour violation (QFV) in the squark sector. The renormalization is done in the $\overline{\rm DR}$ scheme. The gluon and photon radiations are
included by adding the corresponding three-body decay widths. We discuss the dependence of the gluino decay widths on the QFV parameters. The main dependence stems from the $\tilde c_R - \tilde t_R$ mixing in the decays to up-type squarks, and from the $\tilde s_R - \tilde b_R$ mixing in the decays to down-type squarks due to the strong constraints from B-physics on the other quark-flavour mixing parameters. The full one-loop corrections to the gluino decay widths are mostly negative and of the order of about -10\%.
The QFV part stays small in the total width but can vary up to -8\% for the decay width into the lightest squark. For the corresponding branching ratio the effect is somehow washed out by at least a factor of two. The electroweak corrections can be as large as 35\% of the SUSY QCD corrections.
 
%
%

%
\end{abstract}

\clearpage


\section{Introduction}

After the discovery of the Higgs particle in 2012~\cite{HiggsdiscoveryATLAS,HiggsdiscoveryCMS}, a task with high priority of the LHC is the search for new physics, beyond the framework of the Standard Model (SM). One of the most favoured candidates to be discovered are the supersymmetric (SUSY) particles. Their decay chains have been, therefore, extensively studied during the last two decades. Especially relevant are the decays of strongly interacting SUSY particles, squarks and gluinos. At tree-level, the leading gluino decays are those into a quark and a squark. Only when these processes are kinematically forbidden, more-body and loop-induced gluino decays become important.

The decays of the gluino in the Minimal Supersymmetric Standard Model (MSSM) were previously studied with general quark-flavour violation (QFV) in the squark sector at tree-level~\cite{Hurth:2009ke,Bartl:2009au,Bartl:2011wq} or including one-loop corrections with no QFV in the squark sector~\cite{Heinemeyer:2011ab,Beenakker:1996dw}. In this paper we study the two-body decays of the gluino into a scalar quark and a quark at full one-loop level with general quark-flavour mixing in the squark sector of the MSSM. 
Such a study has been performed in detail in~\cite{SebastianPhDth}. The analytical results obtained therein, as well as the developed numerical package FVSFOLD, will be used in the current paper.
Since the experiments on K-physics disfavour mixing between the first two squark generations~\cite{PDG2016}, we only consider mixing between the second and the third generations of squarks. More concrete, we consider scenarios where the gluino only decays into the lightest up- and down-type squarks, $\su_{1,2}$ and $\sd_{1,2}$, which can be mixtures of $\sca_{L,R}$ and $\st_{L,R}$ and $\ss_{L,R}$ and $\sb_{L,R}$, respectively, and all the other decays into $\su_{3,...,6}, \sd_{3,...,6}$ are kinematically forbidden. There exist constraints from B-physics on such mixing as well, which we take into account. The mass limits on SUSY particles as well as the theoretical constraints on the
soft-SUSY breaking trilinear coupling matrices from the vacuum stability conditions are also taken into account. 

In Section~\ref{sec:sq.matrix} we give the formulas for the QFV mixing squark system. In Section~\ref{sec:process} the 
tree-level partial two-body decay widths are derived and then the used $\overline{\rm DR}$ renormalization scheme is explained. 
In order to cure the infrared (IR) divergences, we include the widths of the real gluon/photon radiation process, introducing a small regulator gluon/photon mass. 
In Section~\ref{sec:num} we perform a detailed numerical analysis on the dependences of the two-body decay widths and branching ratios (BRs) on the 
quark flavour-mixing parameters $\durr$ and $\ddrr$ and on the gluino mass.
%
Appendix~\ref{sec:lag} contains the Lagrangian for the gluino-squark-quark interaction. In Appendix~\ref{sec:constr} all constraints we obey are summarized and 
Appendix~\ref{appendix_hardradiation} gives the detailed formulas for the hard radiation of a gluon or a photon.

%
%
\section{QFV parameters in the squark sector of the MSSM}
\label{sec:sq.matrix}

We define the QFV parameters in the up-type squark sector of the MSSM
as follows:
\begin{eqnarray}
\delta^{LL}_{\alpha\beta} & \equiv & M^2_{Q \alpha\beta} / \sqrt{M^2_{Q \alpha\alpha} M^2_{Q \beta\beta}}~,
\label{eq:InsLL}\\[3mm]
\delta^{uRR}_{\alpha\beta} &\equiv& M^2_{U \alpha\beta} / \sqrt{M^2_{U \alpha\alpha} M^2_{U \beta\beta}}~,
\label{eq:InsRR}\\[3mm]
\delta^{uRL}_{\alpha\beta} &\equiv& (v_2/\sqrt{2} ) T_{U\alpha \beta} / \sqrt{M^2_{U \alpha\alpha} M^2_{Q \beta\beta}}~,
\label{eq:InsRL}
\end{eqnarray}
where $\alpha,\beta=1,2,3 ~(\alpha \ne \beta)$ denote the quark flavours $u,c,t$, and $v_{2}=\sqrt{2} \left\langle H^0_{2} \right\rangle$.
Analogously, for the down-type squark sector we have
\begin{eqnarray}
\delta^{dRR}_{\alpha\beta} &\equiv& M^2_{D \alpha\beta} / \sqrt{M^2_{D \alpha\alpha} M^2_{D \beta\beta}}~,
\label{eq:dRR}\\[3mm]
\delta^{dRL}_{\alpha\beta} &\equiv& (v_1/\sqrt{2} ) T_{D\alpha \beta} / \sqrt{M^2_{D \alpha\alpha} M^2_{Q \beta\beta}}~,
\label{eq:dRL}
\end{eqnarray}
where the subscripts $\alpha,\beta=1,2,3 ~(\alpha \ne \beta)$ denote the quark flavours $d,s,b$, and $v_{1}=\sqrt{2} \left\langle H^0_{1} \right\rangle$.
$M_{Q,U,D}$ are the hermitian soft SUSY-breaking squark mass matrices and $T_{U,D}$ are the soft SUSY-breaking trilinear 
coupling matrices of the up- and down-type squarks. These parameters enter the left-left, right-right and left-right blocks of the $6\times6$ squark mass matrix in the super-CKM basis~\cite{Allanach:2008qq},
\begin{equation}
    {\cal M}^2_{\tilde{q}} = \left( \begin{array}{cc}
        {\cal M}^2_{\tilde{q},LL} & {\cal M}^2_{\tilde{q},LR} \\[2mm]
        {\cal M}^2_{\tilde{q},RL} & {\cal M}^2_{\tilde{q},RR} \end{array} \right)\,,
 \label{EqMassMatrix1}
\end{equation}
with $\sq = \su, \sd$. The different blocks in eq.~(\ref{EqMassMatrix1}) are given by
\begin{eqnarray}
    & &{\cal M}^2_{\tilde{u},LL} = V_{\rm CKM} M_Q^2 V_{\rm CKM}^{\dag} + D_{\tilde{u},LL}{\bf 1} + \hat{m}^2_u, \nonumber \\
    & &{\cal M}^2_{\tilde{u},RR} = M_U^2 + D_{\tilde{u},RR}{\bf 1} + \hat{m}^2_u, \nonumber \\
    & & {\cal M}^2_{\tilde{u},RL} = {\cal M}^{2\dag}_{\tilde{u},LR} =
    \frac{v_2}{\sqrt{2}} T_U - \mu^* \hat{m}_u\cot\beta\,,\nn
    & & {\cal M}^2_{\tilde{d},LL} = M_Q^2 + D_{\tilde{d},LL}{\bf 1} + \hat{m}^2_d,  \nonumber \\
    & & {\cal M}^2_{\tilde{d},RR} = M_D^2 + D_{\tilde{d},RR}{\bf 1} + \hat{m}^2_d,\nn
         && {\cal M}^2_{\tilde{d},RL} = {\cal M}^{2\dag}_{\tilde{d},LR} =
   \frac{v_1}{\sqrt{2}} T_D - \mu^* \hat{m}_d\tan\beta,
     \label{RLblocks}
\end{eqnarray}
where
$\mu$ is the higgsino mass parameter, $\tan\beta$ is the ratio of the vacuum expectation values of the neutral Higgs fields $v_2/v_1$, and $\hat{m}_{u,d}$ are the diagonal mass matrices of the up- and down-type quarks.
Furthermore, 
$D_{\tilde{q},LL} = \cos 2\beta m_Z^2 (T_3^q-e_q
\sin^2\theta_W)$ and $D_{\tilde{q},RR} = e_q \sin^2\theta_W \cos 2\beta m_Z^2$,
where
$T_3^q$ and $e_q$ are the isospin and
electric charge of the quarks (squarks), respectively, and $\theta_W$ is the weak mixing
angle. $V_{\rm CKM}$ is the Cabibbo-Kobayashi-Maskawa matrix, which we approximate with the unit matrix.
The squark mass matrix is diagonalized by the $6\times6$ unitary matrices $U^{\tilde{q}}$, such that
\begin{eqnarray}
&&U^{\tilde{q}} {\cal M}^2_{\tilde{q}} (U^{\tilde{q} })^{\dag} = {\rm diag}(m_{\tilde{q}_1}^2,\dots,m_{\tilde{q}_6}^2)\,,
\label{Umatr}
\end{eqnarray}
with $m_{\tilde{q}_1} < \dots < m_{\tilde{q}_6}$, and $\sq=\su, \sd$.
The physical mass eigenstates
$\sq_i, i=1,...,6$ are given by $\sq_i =  U^{\sq}_{i \alpha} \sq_{0\alpha} $.

In this paper we study $\ti{c}_R - \ti{t}_L$, $\ti{c}_L - \ti{t}_R$, $\ti{c}_R - \ti{t}_R$, and $\ti{c}_L - \ti{t}_L$ mixing, 
which is described by the QFV parameters $\delta^{uRL}_{23}$, 
$\delta^{uLR}_{23} \equiv ( \delta^{uRL}_{32})^*$, $\delta^{uRR}_{23}$, and $\dll$, respectively, as well as $\ti{s}_R - \ti{b}_L$, $\ti{s}_L - \ti{b}_R$, $\ti{s}_R - \ti{b}_R$, and $\ti{s}_L - \ti{b}_L$ mixing, 
which is described by the QFV parameters $\delta^{dRL}_{23}$, 
$\delta^{dLR}_{23} \equiv ( \delta^{dRL}_{32})^*$, $\delta^{dRR}_{23}$, and $\dll$, respectively. Note that $\dll$ describes the left-left mixing in both $\su$ and $\sd$ sectors. The
$\ti{t}_R - \ti{t}_L$ mixing is described by the quark-flavour conserving (QFC) parameter $\delta^{uRL}_{33}$.  All parameters mentioned are assumed to be real.

\section{Two-body decays of gluino at full one-loop level in the general MSSM}
\label{sec:process}
%
We study two-body decays of gluino into a squark and a quark, $\sg \to \sq^* q$.
The tree-level partial decay widths $\Gamma^0(\sg \to \sq^*_i q_g)$, with $i=1,...,6$, $q = u, d$, and the subscript $g$ is the quark-generation index, is given by
\bea
\Gamma^0(\sg \to \sq^*_i q_g) = {c\, \lambda^{1/2}(\msg^2,m_{\sq_i}^2, m_{q_g}^2) \over 64\, \pi\,  \msg^3}|{\cal M}_0|^2\,, 
\label{decaywidttree}
\eea
where $c = 1/16$ is the average factor for  the incoming $\sg$.
The tree-level amplitude squared reads
\be
|{\cal M}_0|^2 = (|g^i_L|^2+|g^i_R|^2)(\msg^2-m_{\sq_i}^2 +m_{q_g}^2) + 2 \,\msg m_{q_g}(g^{i*}_L g^i_R+g^i_L g_R^{i*})\,,
\label{mat2tree}
\ee
with $\lambda(x^2,y^2, z^2) = x^2+y^2+z^2-2xy -2xz-2yz$, no summation over $i$, and the tree-level couplings $g^i_{L,R}$ are given by (see also Appendix~{\ref{sec:lag}}) 
\be
g^i_L=-\sqrt{2}~g_s T U^{\sq}_{i,g}\,, \quad g^i_R=\sqrt{2}~g_s T U^{\sq}_{i,g+3}\,,
 \label{treecoup}
\ee
where T are the generators of the the SU(3) colour group, and $U^{\sq}$, with $\sq=\su, \sd$ are the up- and down-squark mixing matrices defined by eq.~(\ref{Umatr}).
By inserting eq.~(\ref{treecoup}) into eq.~(\ref{mat2tree}) and using tr$(T^a T^a) = N_c C_F = 4$ we can write eq.~(\ref{decaywidttree}) in the 
explicit form 
\bea
\Gamma^0(\sg \to \sq^*_i q_g) & = & {\lambda^{1/2}(\msg^2,m_{\sq_i}^2, m_{q_g}^2) \over 32\,  \msg^3} \alpha_s\, \bigg(
\left(|U^{\sq}_{i,g}|^2+|U^{\sq}_{i,g+3}|^2\right)(\msg^2-m_{\sq_i}^2 +m_{q_g}^2) \nn 
&& \hspace*{5cm}- 4 \msg m_{q_g}{\rm Re}\left(U^{\sq *}_{i,g}\, U^{\sq}_{i,g+3} \right) \bigg)\, .
\label{decaywidttree}
\eea
%
In order to obtain an ultraviolet (UV) convergent result at one-loop level we employ the dimensional reduction ($\overline {\rm DR}$) regularisation scheme, which implies that all tree-level input parameters of the Lagrangian are defined at the scale $Q=M_3\approx m_{\sg}$. Since in this scheme the tree-level couplings 
$g^i_{L,R}$ are defined at the scale $Q$, they do not receive further finite shifts due to radiative corrections. The physical scale independent masses and fields are obtained from the $\overline {\rm DR}$ ones using on-shell renormalisation conditions. \\

\noindent
We can write the renormalised one-loop partial decay widths as
\bea
\label{decaywidth}
\hspace*{-2cm}
\Gamma(\sg \to \sq^*_i q_g) &=& \Gamma^0( \sg \to \sq^*_i q_g)~+~\Delta \Gamma(\sg \to \sq^*_i q_g)\, , \quad {\rm with}  \\
\Delta \Gamma(\sg \to \sq^*_i q_g)  &= & {c\,\lambda^{1/2}(\msg^2,m_{\sq_i}^2, m_{q_g}^2) \over 32\, \pi\,  \msg^3} {\rm Re}({\cal M}_0^\dagger {\cal M}_1)\,,
\quad {\rm and} \nn
 {\rm Re}({\cal M}_0^\dagger {\cal M}_1) & = & {\rm Re}\bigg( (g^{i*}_L \Delta g_L + g^{i*}_R \Delta g_R)(\msg^2-m_{\sq_i}^2 +m_{q_g}^2) \nn
 && \hspace*{4cm} + 2 \msg m_{q_g}(g^{i*}_L \Delta g_R +  g^{i*}_R \Delta g_L)\bigg) \nonumber\, ,
\eea
where $ {\cal M}_1$ is the one-loop amplitude. The complete list of diagrams can be found in the Appendix of~\cite{SebastianPhDth}. 
The one-loop shifts to the coupling constants, $\Delta g_L$ and $\Delta g_R$, receive contributions from all vertex diagrams, the amplitudes arising from the wave-function renormalisation constants and the amplitudes arising from the coupling counter-terms~\footnote{Note, that in the $\overline {\rm DR}$ scheme the coupling corrections contain only UV divergent terms which have to be canceled exactly to yield a convergent result.},
\be
\Delta g_{L,R} = \d g_{L,R}^v + \d g_{L,R}^w +\d g_{L,R}^c\,,
\ee
where $\d g_{L,R}^v$ is due to all vertex radiative corrections, and $\d g_{L,R}^c$ is due to the coupling counter terms.
The wave-function induced corrections are given by
\bea
\d g_{L}^{w, {\rm diag}} &=& {1\over 2} \left( \d Z^{\sg R*} + \d Z_{ii}^{\sq*} +\d Z_{gg}^{q L} \right) g_L^i\,,\nn
\d g_{R}^{w,  {\rm diag}} &=& {1\over 2} \left( \d Z^{\sg L*} + \d Z_{ii}^{\sq*} +\d Z_{gg}^{q R} \right) g_R^i\,, \nn
\d g_{L}^{w,  {\rm off-diag}} &=& {1\over 2} \left( \d Z_{ij}^{\sq*} g_L^j +\d Z_{lg}^{q L}  g_L^{i,l} \right)\,,\nn
\d g_{R}^{w,  {\rm off-diag}} &=& {1\over 2} \left( \d Z_{ij}^{\sq*} g_R^j +\d Z_{lg}^{q R} g_R^{i,l} \right)\,,
\label{renconsts}
\eea
with $i$ and $j$ fixed, $j \neq i$, $l \neq j$. Note that $g_{L,R}^{i,l}$ denote the $\sg \sq^{*}_i \bar q_l$ couplings.
The explicit expressions for the renormalisation constants $\d Z$ in (\ref{renconsts}) can be found in~\cite{SebastianPhDth}.

To cure the infrared (IR) divergences, in addition to (\ref{decaywidth}), we include the widths of the real gluon/photon radiation processes, $\Gamma(\sg \to \sq_i q_g g /\gamma)$, assuming a small regulator gluon/photon mass $\lambda$. The explicit formulas for the hard radiation widths are given in Appendix~{\ref{appendix_hardradiation}.

The full one-loop contribution to the total two-body decay width, see (\ref{decaywidth}), is due to SUSY-QCD and electroweak corrections,
\be
\Gamma(\sg \to \sq^* q) = \Gamma^0(\sg \to \sq^* q)+\Delta \Gamma^{\rm SQCD}(\sg \to \sq^* q) + \Delta \Gamma^{\rm EW}(\sg \to \sq^* q)\,.
\ee
$\Delta \Gamma^{\rm SQCD}$ includes loops with gluon and gluino, and $\Delta \Gamma^{\rm EW}$ includes
loops with EW gauge bosons, photon, Higgs bosons and EWinos.   
In the numerical analyses performed in~\cite{SebastianPhDth}, as well as in~\cite{Heinemeyer:2011ab}, it was shown that in the considered scenarios the electroweak corrections are not negligible, but necessary for a correct one-loop evaluation. 
As you will see, in our numerical analysis we will come to a similar conclusion.

Hereafter we will use the notation $\Gamma(\sg \to \sq_i q_g) = \Gamma(\sg \to \sq^*_i q_g) + \Gamma(\sg \to \sq_i \bar q_g)$. In our case where CP is conserved this is equivalent with $2 \Gamma(\sg \to \sq^*_i q_g)$.

\section{Numerical results}
\label{sec:num}

In order to demonstrate quantitatively our results on the gluino 
decay widths and branching ratios we first fix a 
reference scenario and then vary the QFV parameters within the allowed region. Our reference scenario fulfils all relevant theoretical and experimental constraints, which we 
discuss in more detail in Appendix~\ref{sec:constr}. 
The input parameters and the physical output parameters are shown in Tables~\ref{basicparam} and \ref{physmasses},
respectively. The flavour decomposition of the $\su_{1,2}$ and $\sd_{1,2}$ squarks is shown in Table~\ref{flavourdecomp}. 
For calculating the $h^0$ mass and the low-energy observables, especially 
those ones in the B-sector (see Table~\ref{TabConstraints}), we use the public code 
SPheno v3.3.3~\cite{SPheno1, SPheno2}. 
The gluino two-body widths and branching ratios at full one-loop level in the MSSM with QFV are calculated with the numerical code FVSFOLD, developed in~\cite{SebastianPhDth}. For building FVSFOLD the packages FeynArts~\cite{Hahn:2000kx,Hahn} and FormCalc~\cite{Hahn:1998yk}\
were used. Furthermore, we use LoopTools~\cite{Hahn:1998yk} based on the FF package~\cite{FF}, and SSP~\cite{SSP}.
In order to have simultanously 
a UV and IR finite result we first calculate the total result 
by using only ${\overline {\rm DR}}$ parameters for the 
one-loop partial width including the real hard radiation. 
Then we use on-shell masses, which are calculated within FVSFOLD, 
in the kinematic two-body prefactor $\lambda^{1/2}/\msg^3$, see 
eq.~(\ref{decaywidttree}).\\

The scenario shown in Table~\ref{basicparam} violates quark-flavour explicitly in both up- and down-squark sectors. The values of the parameters $M_{1,2,3}$ are chosen to satisfy approximately the GUT relations ($M_1:M_2:M_3=1:2:6$). The Higgs boson $h^0$ is SM-like with $m_{h^0}=125~\gev$ and all other Higgses are much heavier in mass and degenerate.  The ratio of the vacuum expectation values of the neutral Higgs fields $v_2/v_1$ is taken relatively small, $\tan \b = 15$. The value of the $\mu$ parameter is also chosen small for naturalness reasons. The flavour decompositions of the $\su_{1,2}$ and $\sd_{1,2}$ squarks are shown in Table~\ref{flavourdecomp}. In this scenario the $\su_1$ squark is a strong mixture of $\sca_R$ and $\st_R$, with a tiny contribution from $\sca_L$, and the $\su_2$ squark is mainly $\st_L$, with a tiny contribution from $\sca_R$. The $\sd_1$ is a mixture of $\ss_R$ and $\sb_R$, and $\sd_2$ is a pure  $\sb_L$.\\

\begin{table}[h!]
\caption{QFV reference scenario: all parameters are calculated
at $Q = M_3 = 3~{\rm TeV} \simeq m_{\sg}$,
except for $m_{A^0}$ which is the pole mass of $A^0$, 
and $T_{U33} =2500$~GeV (corresponding to $\delta^{uRL}_{33} = 0.06$). All other squark parameters are zero.}
\begin{center}
\begin{tabular}{|c|c|c|c|c|c|}
  \hline
 $M_1$ & $M_2$ & $M_3$ &$\mu$ & $\tan \beta$ & $m_{A^0}$ \\
 \hline \hline
 500~\gev  &  1000~\gev & 3000~\gev &500~\gev & 15 &  3000~\gev \\
  \hline
\end{tabular}
\vskip0.4cm
\begin{tabular}{|c|c|c|c|}
  \hline
   & $\alpha = 1$ & $\alpha= 2$ & $\alpha = 3$ \\
  \hline \hline
   $M_{Q \alpha \alpha}^2$ & $3200^2~\gev^2$ &  $3000^2~\gev^2$  & $2600^2~\gev^2$ \\
   \hline
   $M_{U \alpha \alpha}^2$ & $3200^2~\gev^2$ & $3000^2~\gev^2$ & $2600^2~\gev^2$ \\
   \hline
   $M_{D \alpha \alpha}^2$ & $3200^2~\gev^2$ & $3000^2~\gev^2$ &  $2600^2~\gev^2$  \\
   \hline
\end{tabular}
\vskip0.4cm
\begin{tabular}{|c|c|c|c|c|c|c|}
  \hline
   $\delta^{LL}_{23}$ & $\delta^{uRR}_{23}$  &  $\delta^{uRL}_{23}$ & $\delta^{uLR}_{23}$ &$\delta^{dRR}_{23}$  &  $\delta^{dRL}_{23}$ & $\delta^{dLR}_{23}$\\
  \hline \hline
    0.01 & 0.7 & 0.04  & 0.07 & 0.7 & 0  & 0   \\
    \hline
\end{tabular}
\end{center}
\label{basicparam}
\end{table}
%
\begin{table}[h!]
\caption{Physical masses of the particles in GeV for the scenario of Table~\ref{basicparam}.}
\begin{center}
\begin{tabular}{|c|c|c|c|c|c|}
  \hline
  $\mnt{1}$ & $\mnt{2}$ & $\mnt{3}$ & $\mnt{4}$ & $\mch{1}$ & $\mch{2}$ \\
  \hline \hline
  $460$ & $500$ & $526$ & $1049$ & $493$ & $1049$ \\
  \hline
\end{tabular}
\vskip 0.4cm
\begin{tabular}{|c|c|c|c|c|}
  \hline
  $m_{h^0}$ & $m_{H^0}$ & $m_{A^0}$ & $m_{H^+}$ \\
  \hline \hline
  $125$  & $3000$ & $3000$ & $3001$ \\
  \hline
\end{tabular}
\vskip 0.4cm
\begin{tabular}{|c|c|c|c|c|c|c|}
  \hline
  $\msg$ & $\msu{1}$ & $\msu{2}$ & $\msu{3}$ & $\msu{4}$ & $\msu{5}$ & $\msu{6}$ \\
  \hline \hline
  $3154$ & $1602$ & $2686$ & $3087$ & $3295$ & $3300$ & $3692$ \\
  \hline
\end{tabular}
\vskip 0.4cm
\begin{tabular}{|c|c|c|c|c|c|}
  \hline
   $\msd{1}$ & $\msd{2}$ & $\msd{3}$ & $\msd{4}$ & $\msd{5}$ & $\msd{6}$ \\
  \hline \hline
  $1662$ & $2689$ & $3087$ & $3295$ & $3301$ & $3747$ \\
  \hline
\end{tabular}
\end{center}
\label{physmasses}
\end{table}
%
\begin{table}[h!]
\caption{Flavour decomposition of $\su_{1,2}$ and $\sd_{1,2}$ for the scenario of Table~\ref{basicparam}. Shown are the squared coefficients. }
\begin{center}
\begin{tabular}{|c|c|c|c|c|c|c|c|}
  \hline
  & $\su_L$ & $\sca_L$ & $\st_L$ & $\su_R$ & $\sca_R$ & $\st_R$ \\
  \hline \hline
 $\su_1$  & $0$ & $0.004$ & $0$ & $0$ & $0.38$ & $0.61$ \\
  \hline 
  $\su_2$  & $0$ & $0.001$ & $0.99$ & $0$ & $0.006$ & $0$ \\
  \hline
\end{tabular}
\vskip 0.4cm
\begin{tabular}{|c|c|c|c|c|c|c|c|}
  \hline
  & $\sd_L$ & $\ss_L$ & $\sb_L$ & $\sd_R$ & $\ss_R$ & $\sb_R$ \\
  \hline \hline
 $\sd_1$  & $0$ & $0$ & $0$ & $0$ & $0.4$ & $0.6$ \\
  \hline 
  $\sd_2$  & $0$ & $0$ & $1$ & $0$ & $0$ & $0$ \\
  \hline
\end{tabular}
\end{center}
\label{flavourdecomp}
\end{table}
At our reference parameter point the gluino decays into $\su_{1,2}\,c$, $\su_{1,2} \,t$, $\sd_{1,2} \,s$ and $\su_{1,2} \,b$ are kinematically allowed, with branching ratios B$(\sg \to \su_1\,c)\approx 17\%$, B$(\sg \to \sd_1\,s)\approx 18\%$, B$(\sg \to \su_1\,t) = {\rm B}(\sg \to \sd_1\,b) \approx 27\%$, B$(\sg \to \su_2\,t)\approx 5\%$ and the others being very small. The total two-body width including the full one-loop contribution, $\Gamma(\sg \to \sq q) = 70\,\gev$, where the tree-level width $\Gamma^0(\sg \to \sq q) = 75\,\gev$\footnote{A comparison with the tree-level results is not 
precisely accurate, since in the $\overline{\rm DR}$ scheme 
the tree-level width alone does not have a physical meaning, 
but the width at full one-loop level does. However, in order 
to get approximately an idea how large the tree-level result 
is, we allow to use on-shell masses only in the kinematics 
factor of eq.~(\ref{decaywidttree}). In the following, we will 
call this tree-level result.}, and the SUSY-QCD and the electroweak corrections are negative, $\Delta \Gamma^{\rm SQCD}(\sg \to \sq q) = -4.6\,\gev$ and  $\Delta \Gamma^{\rm EW}(\sg \to \sq q) = -0.5\,\gev$, giving about -6.4\% and -0.7\% of the total two-body gluino width $\Gamma(\sg \to \sq q)$, respectively. 
Note, that the SQCD contribution includes gluon and gluino, and the EW contribution includes also the photon.
In the same scenario with no 
quark-flavour violation, i.e. when all QFV ($\d$) parameters 
listed in Table~\ref{basicparam} are set to zero, 
we have $\Gamma(\sg \to \sq q) = 14\,\gev$.\\

\begin{figure*}[h!]
\centering
\subfigure[]{
   { \mbox{\hspace*{-1cm} \resizebox{7.5cm}{!}{\includegraphics{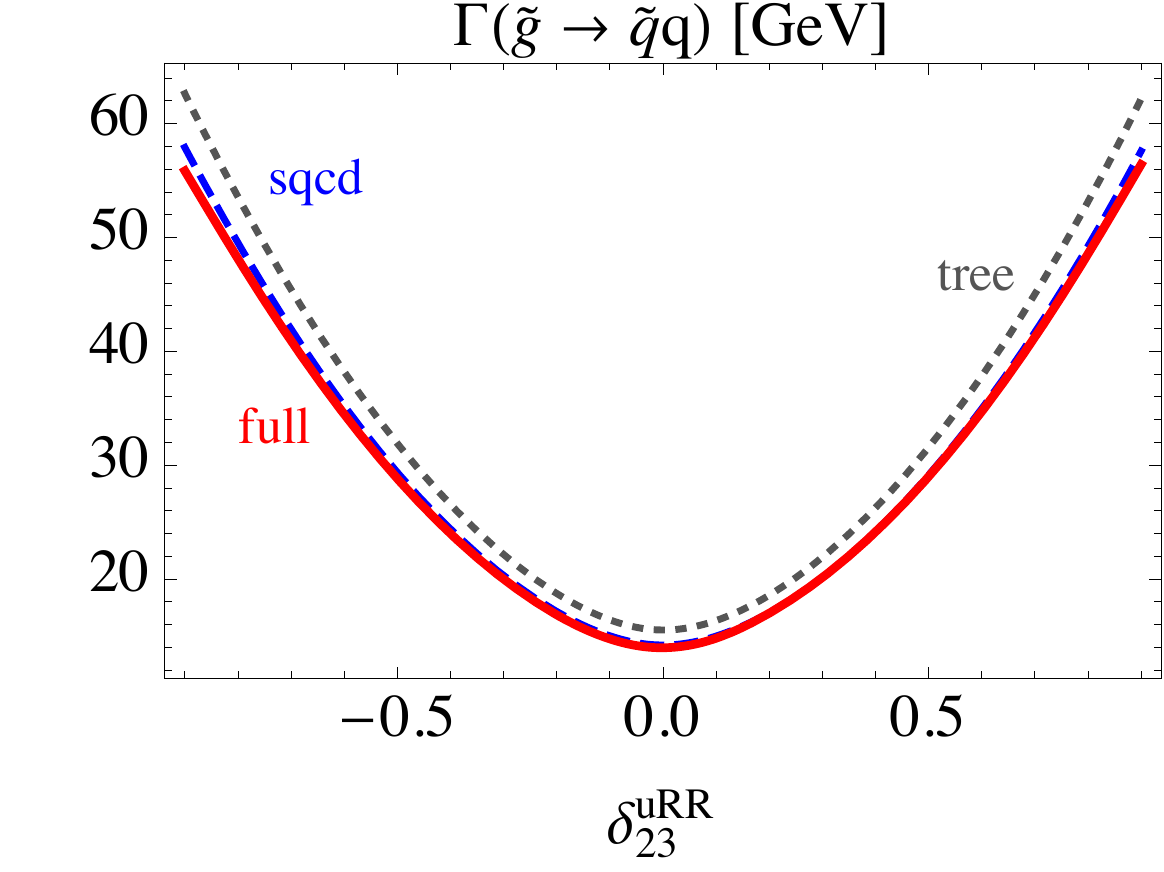}} \hspace*{-0.8cm}}}
   \label{fig1a}}
\subfigure[]{
   { \mbox{\hspace*{0.5cm} \resizebox{8cm}{!}{\includegraphics{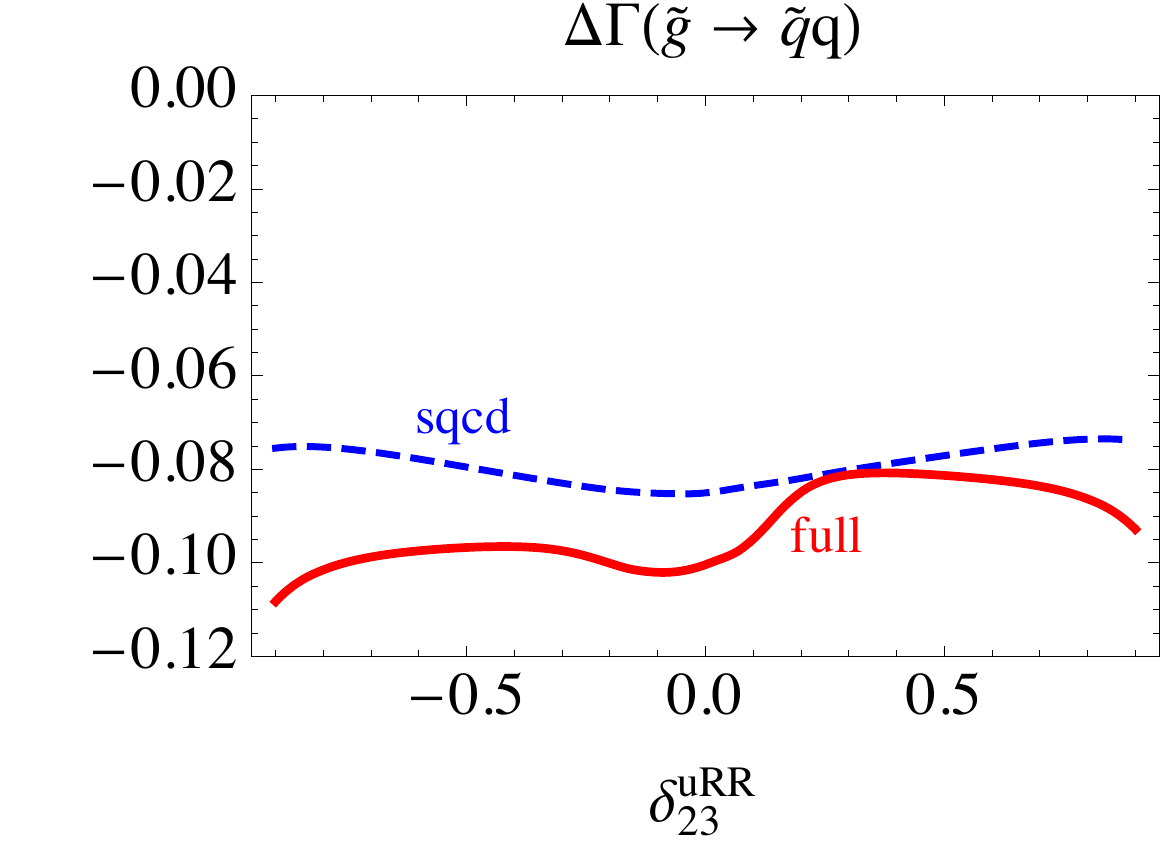}} \hspace*{-0.8cm}}}
   \label{fig1b}}\\
\subfigure[]{
   { \mbox{\hspace*{-1cm} \resizebox{7.5cm}{!}{\includegraphics{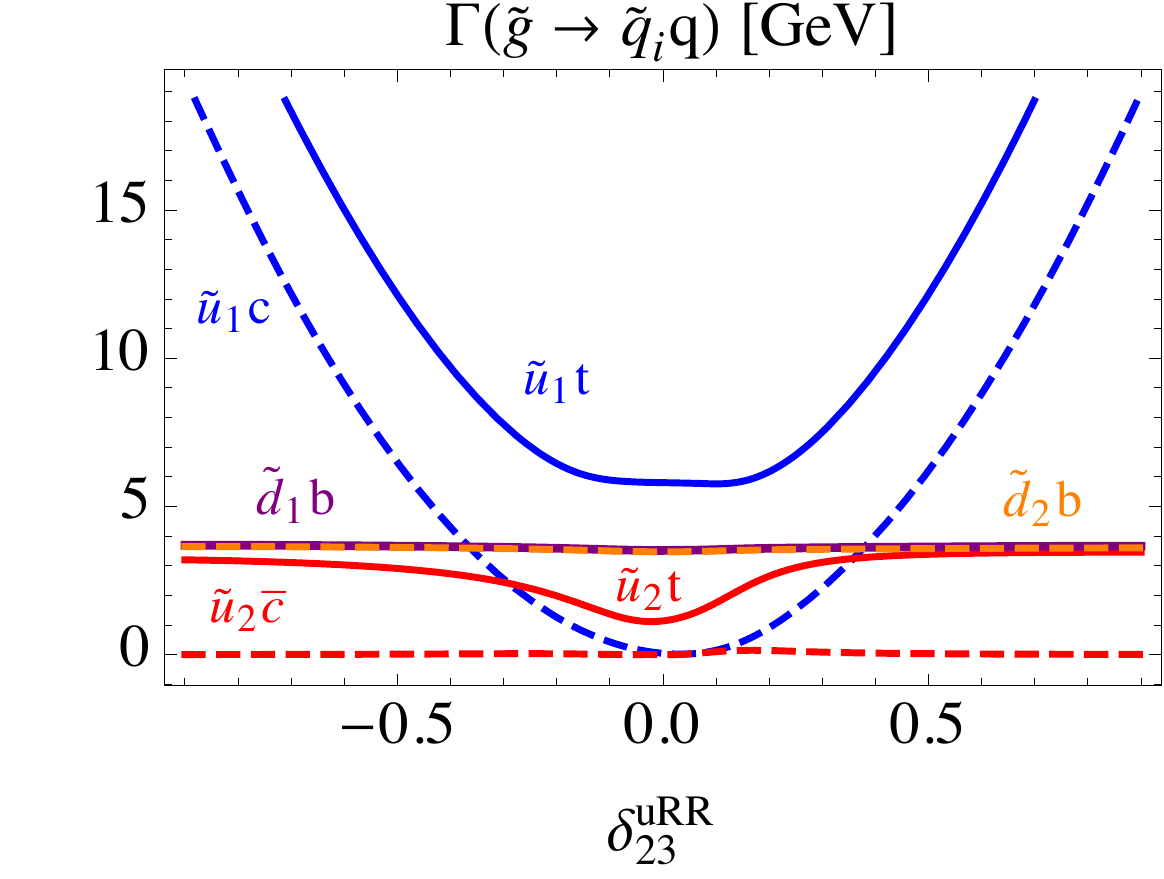}} \hspace*{-0.8cm}}}
   \label{fig1c}}
 \subfigure[]{
   { \mbox{\hspace*{+0.5cm} \resizebox{7.5cm}{!}{\includegraphics{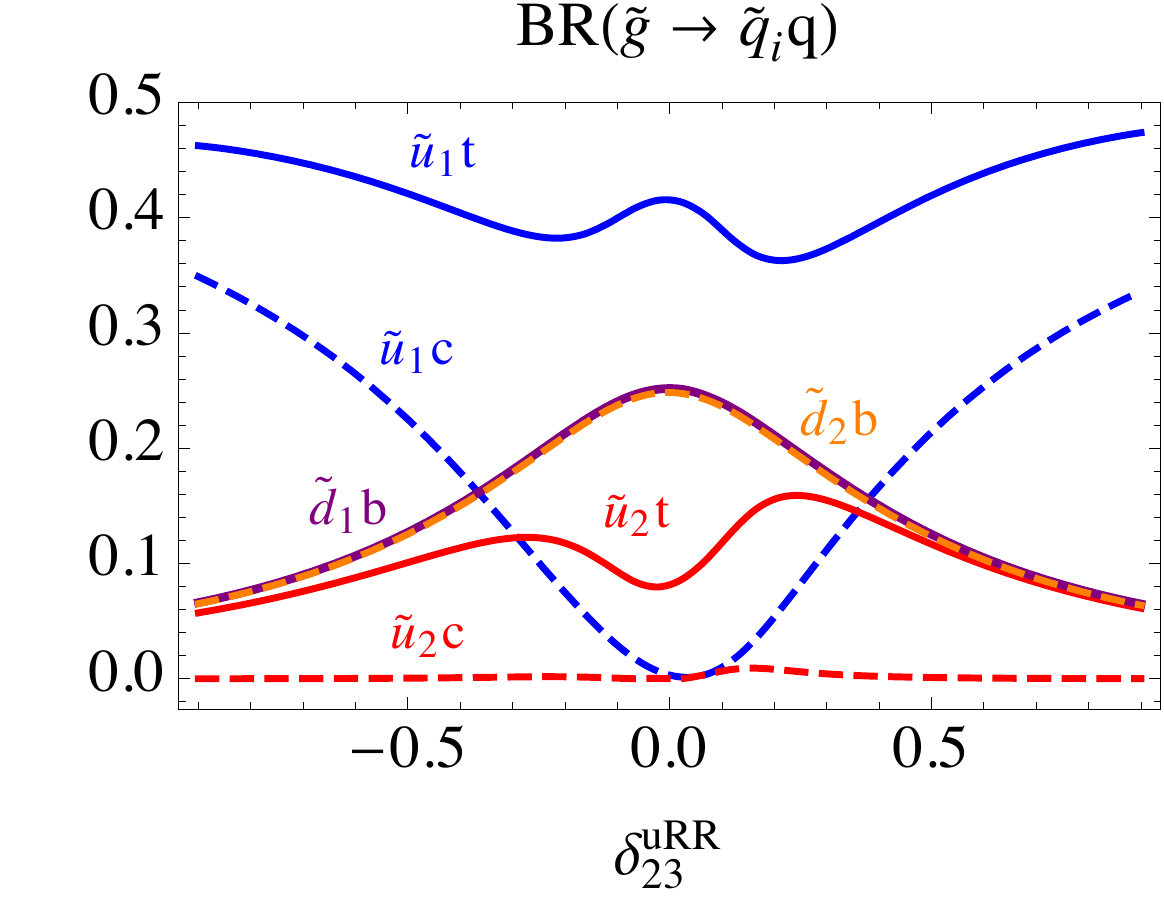}} \hspace*{-1cm}}}
  \label{fig1d}}
\caption{(a) Total two-body decay width $\Gamma(\sg \to \sq q)$
at tree-level, SQCD one-loop and full one-loop corrected as
functions of the QFV parameter $\durr$;
(b) $\Delta \Gamma(\sg \to \sq q)$ being the SQCD one-loop and the
full one-loop corrections to $\Gamma(\sg \to \sq q)$ relative to
the tree-level width; (c) Partial decay widths and (d) branching
ratios of the kinematically allowed individual two-body channels
at full one-loop level as functions of $\durr$.
All the other parameters are fixed as in Table~\ref{basicparam},
except $\durl = \dulr = 0.03$.}
\label{fig1}
\end{figure*}
The QFV left-right mixing, described by the parameters 
$\dulr, \durl, \ddlr, \ddrl$, is constrained from the 
vacuum stability conditions (see Section~\ref{sec:constr}) 
and is required to be rather small. On another hand, a 
sizable value of $\dll$ is not possible because it violates 
B-physics constraints such as the B($B_s \to \mu^+ \mu^-$) 
constraint. However, large right-right mixing in both 
$\su$ and $\sd$ sectors is allowed and therefore, in the 
following, we show only plots with dependences on the 
$\durr$ and $\ddrr$ parameters.\\

In Fig.~\ref{fig1} we show dependences on the QFV parameter $\durr$. In~\ref{fig1a} 
the tree-level, the SQCD and total full one-loop widths 
and in~\ref{fig1b} the relative contributions of the one-loop SQCD and the full one-loop part in terms of the tree-level result are shown. The partial decay widths as well as the branching ratios of the kinematically allowed two-body channels at full one-loop level are shown in~\ref{fig1c} and~\ref{fig1d}, respectively.
In \ref{fig1a} it is seen that $\Gamma(\sg \to \sq q)$ is quite sensitive to the parameter $\durr$. 
The dependence of the tree-level width and the full one-loop corrected width is similar and their difference becomes a little more important for large absolute values of $\durr$. 
This means that the QFV parameter dependence is mainly due to the kinematic factor, see Section~\ref{sec:process}.  
The SQCD correction shown in~\ref{fig1b} is only weakly dependent on 
$\durr$ and is about -8\%. The EW correction can become -3\% for large and negative values of $\durr$. In Fig.~\ref{fig1c} the partial widths of the $\sd_{1,2} b$ modes coincide because $m_{\sd_1}\approx m_{\sd_2}$.
The same holds for the branching ratios in~\ref{fig1d} . For $\durr \approx 0$ the width of $\tilde g \to \tilde u_1 c$ becomes tiny because then $\tilde u_1$ is mainly $\tilde t_R$ as all the other QFV $\delta$'s are relatively small.\\

\begin{figure*}[h!]
\centering
\subfigure[]{
   { \mbox{\hspace*{-1cm} \resizebox{8cm}{!}{\includegraphics{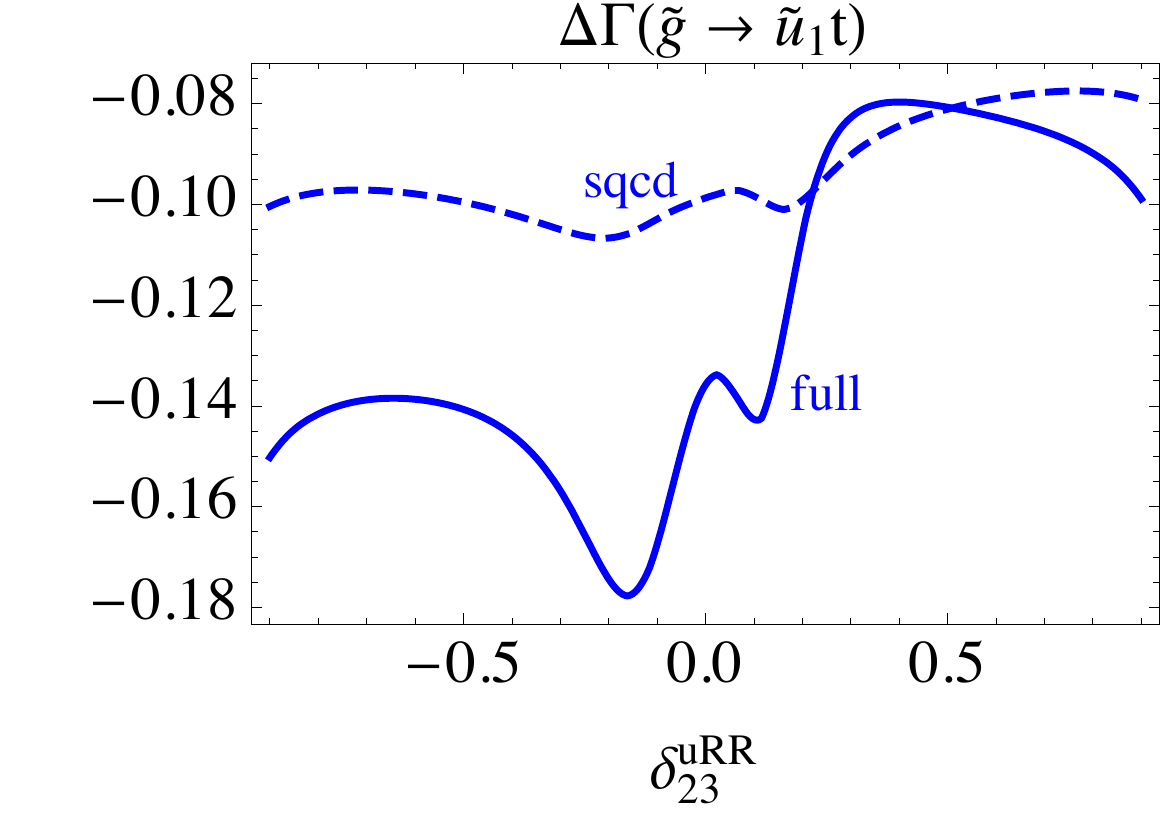}} \hspace*{-0.8cm}}}
   \label{fig1Da}}
\subfigure[]{
   { \mbox{\hspace*{0.5cm} \resizebox{8cm}{!}{\includegraphics{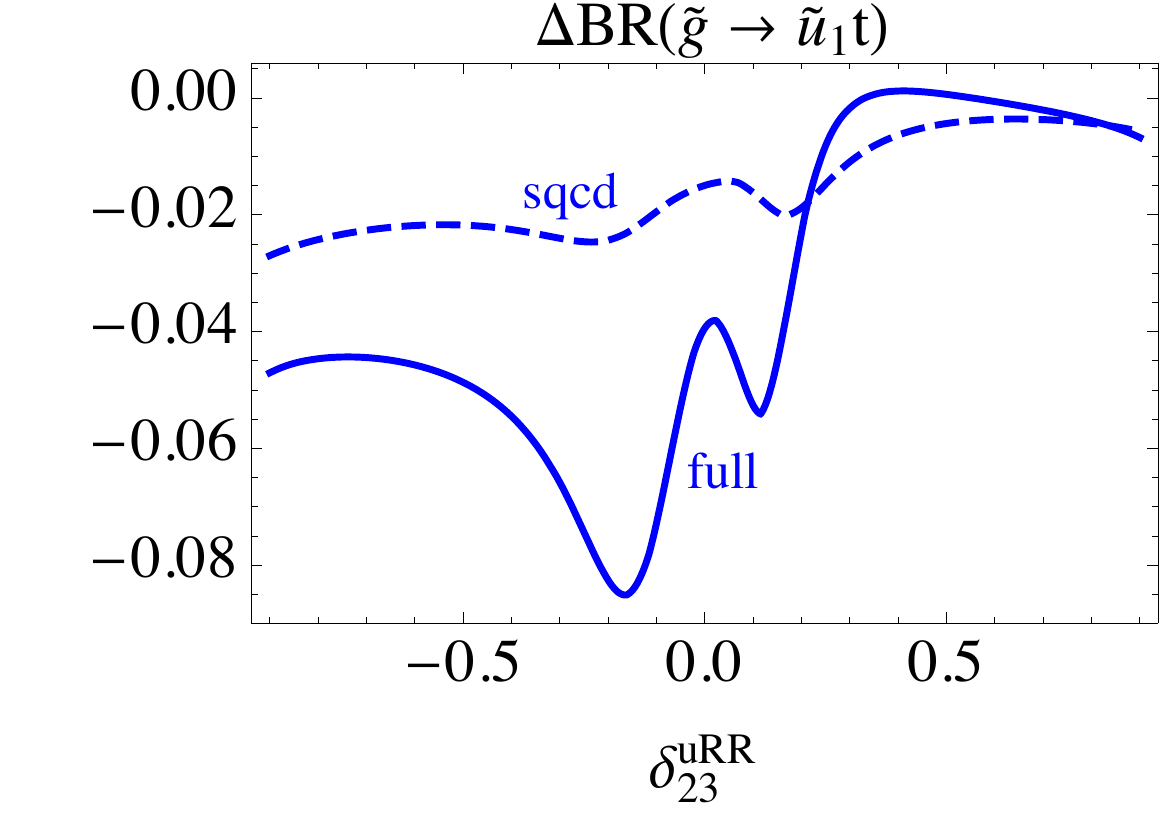}} \hspace*{-0.8cm}}}
   \label{fig1Db}} 
\caption{$\Delta \Gamma$ and $\Delta$BR denote the SQCD
one-loop and the full one-loop corrections relative to
the tree-level result for the decay $\sg \to \su_1 t$
as a function of $\durr$; (a) and (b) is for the partial
width and the branching ratio, respectively.
The other parameters are fixed as in Fig.~\ref{fig1}.}
\label{fig1D}
\end{figure*}
%
Fig.~\ref{fig1D} shows the relative contribution of the one-loop SQCD and the full one-loop part in terms of the tree-level result for the 
partial decay width~\ref{fig1Da} and the branching ratio~\ref{fig1Db} of the decay $\tilde g \to \tilde u_1 t$ as a function of $\durr$. We see in~\ref{fig1Da}  
that the SQCD corrections  
vary in the range of -8\% to - 10\%. The EW correction is much stronger dependent on $\durr$ varying between 1\% down to -8\%. 
The effects are similar in the branching ratio (b), but weaker.
Out of the squark masses only $m_{\tilde u_1}$ is strongly dependent on $\durr$. 
In the whole range of $\durr$ no additional channel opens but those visible in Figs.~\ref{fig1c} and~\ref{fig1d} . Therefore, the wiggles stem from the complex 
structures of the QFV one-loop contributions.\\

%
\begin{figure*}[h!]
\centering
\subfigure[]{
   { \mbox{\hspace*{-1cm} \resizebox{7.5cm}{!}{\includegraphics{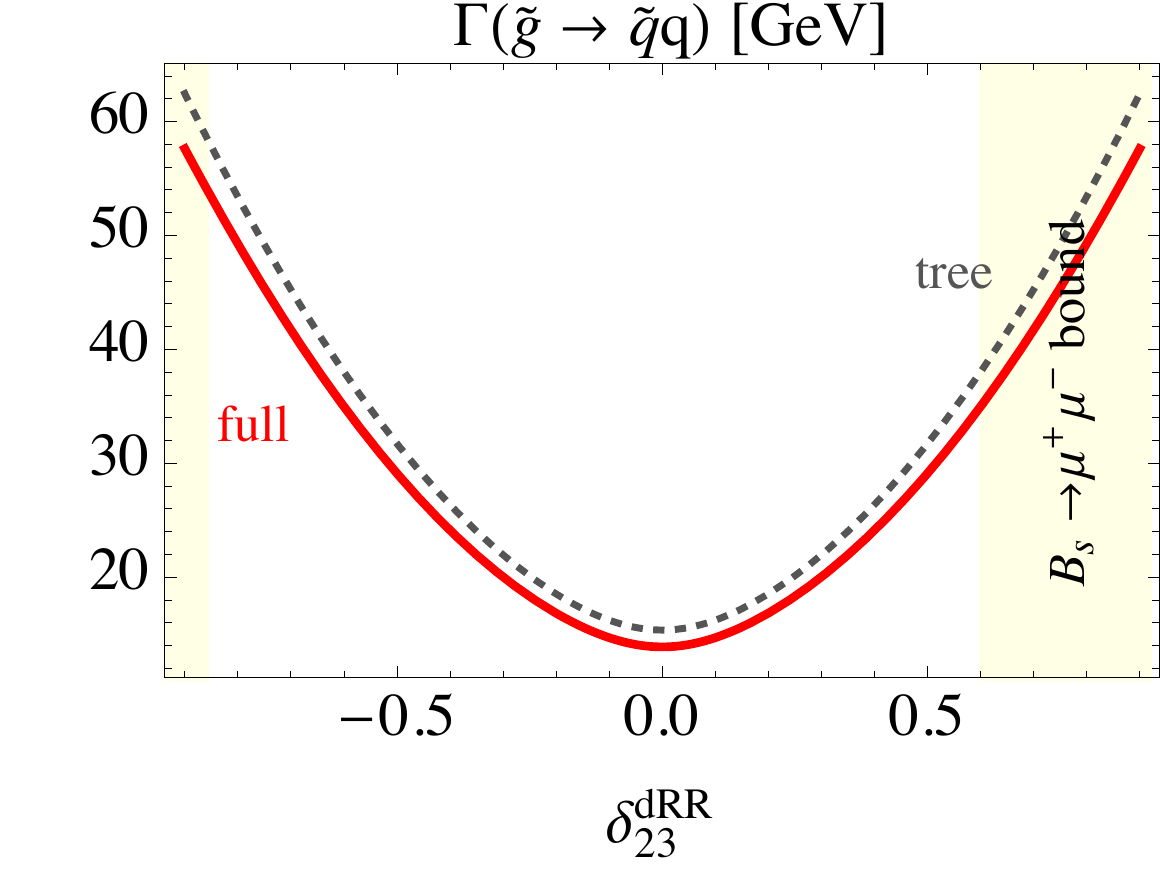}} \hspace*{-0.8cm}}}
   \label{fig2a}}
\subfigure[]{
   { \mbox{\hspace*{0.5cm}\resizebox{8cm}{!}{\includegraphics{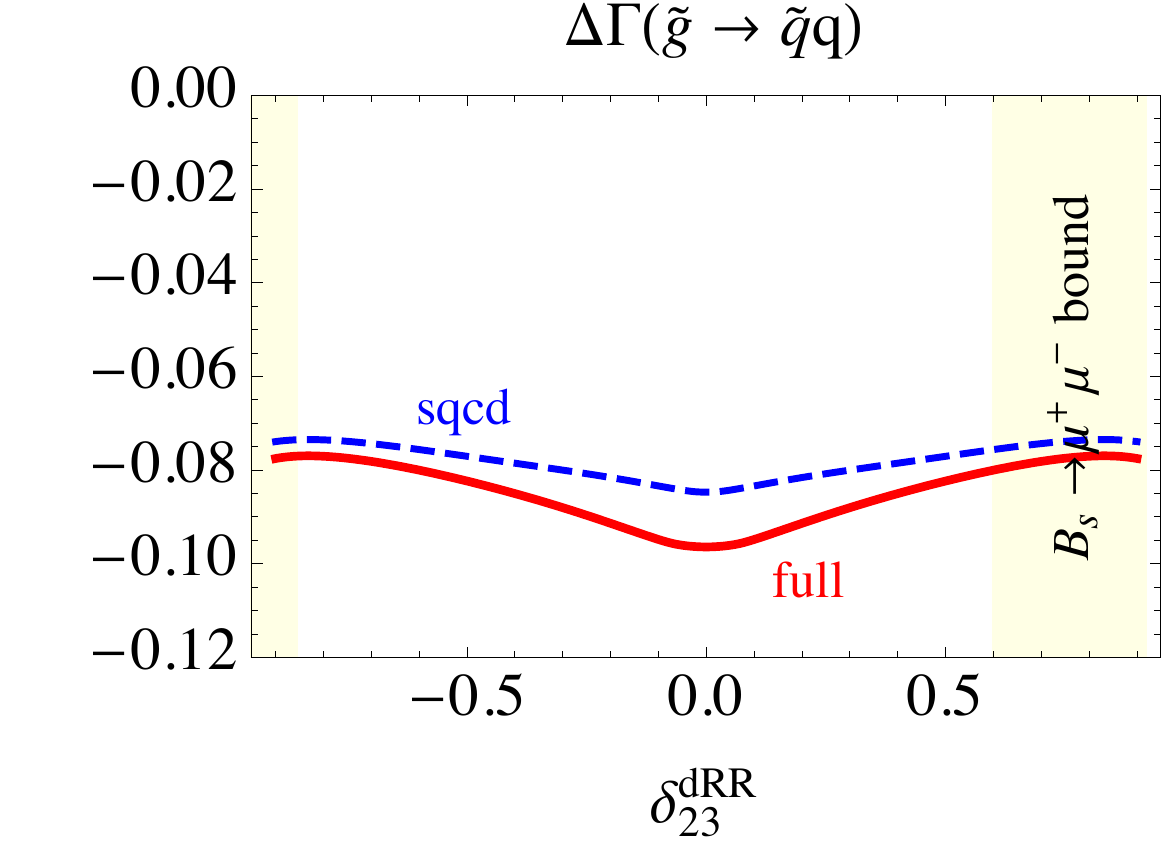}} \hspace*{-0.8cm}}}
   \label{fig2b}}\\
\subfigure[]{
   { \mbox{\hspace*{-1cm} \resizebox{7.5cm}{!}{\includegraphics{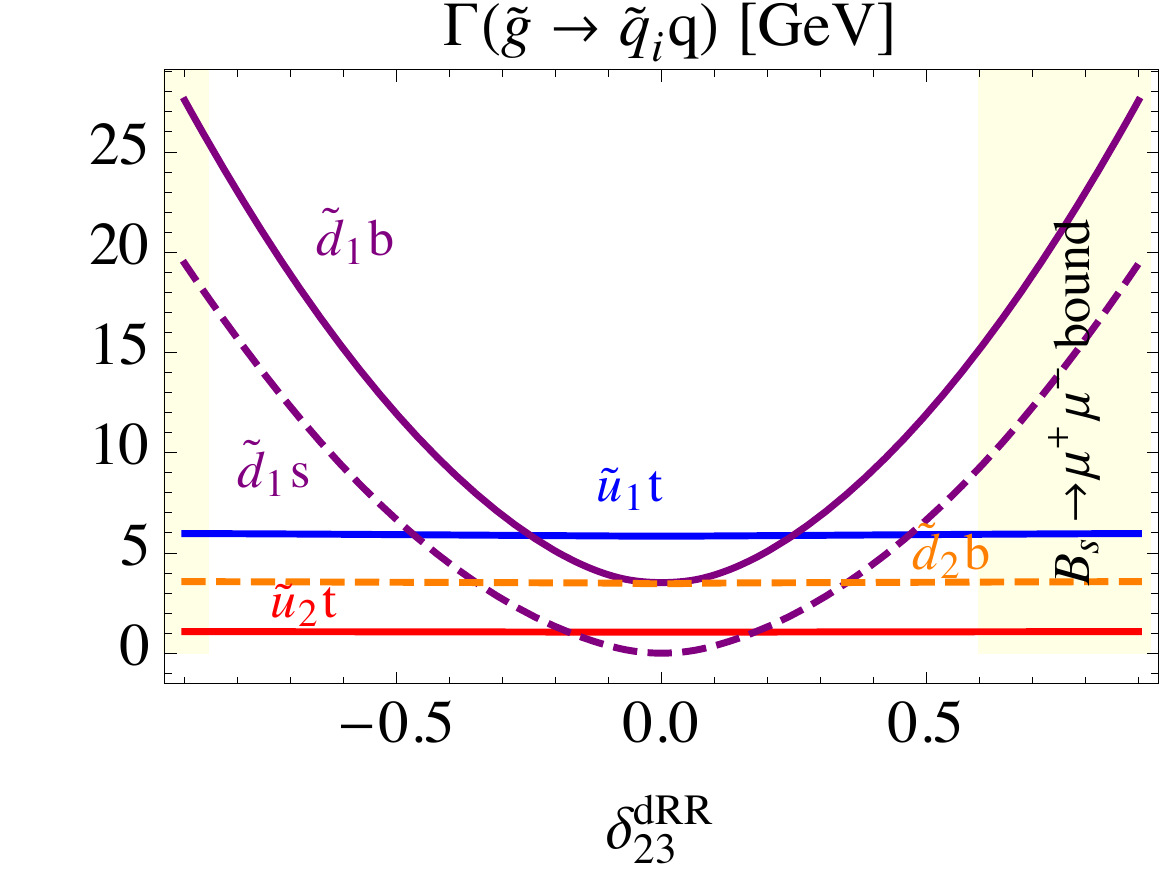}} \hspace*{-0.8cm}}}
   \label{fig2c}}
 \subfigure[]{
   { \mbox{\hspace*{+0.5cm} \resizebox{7.5cm}{!}{\includegraphics{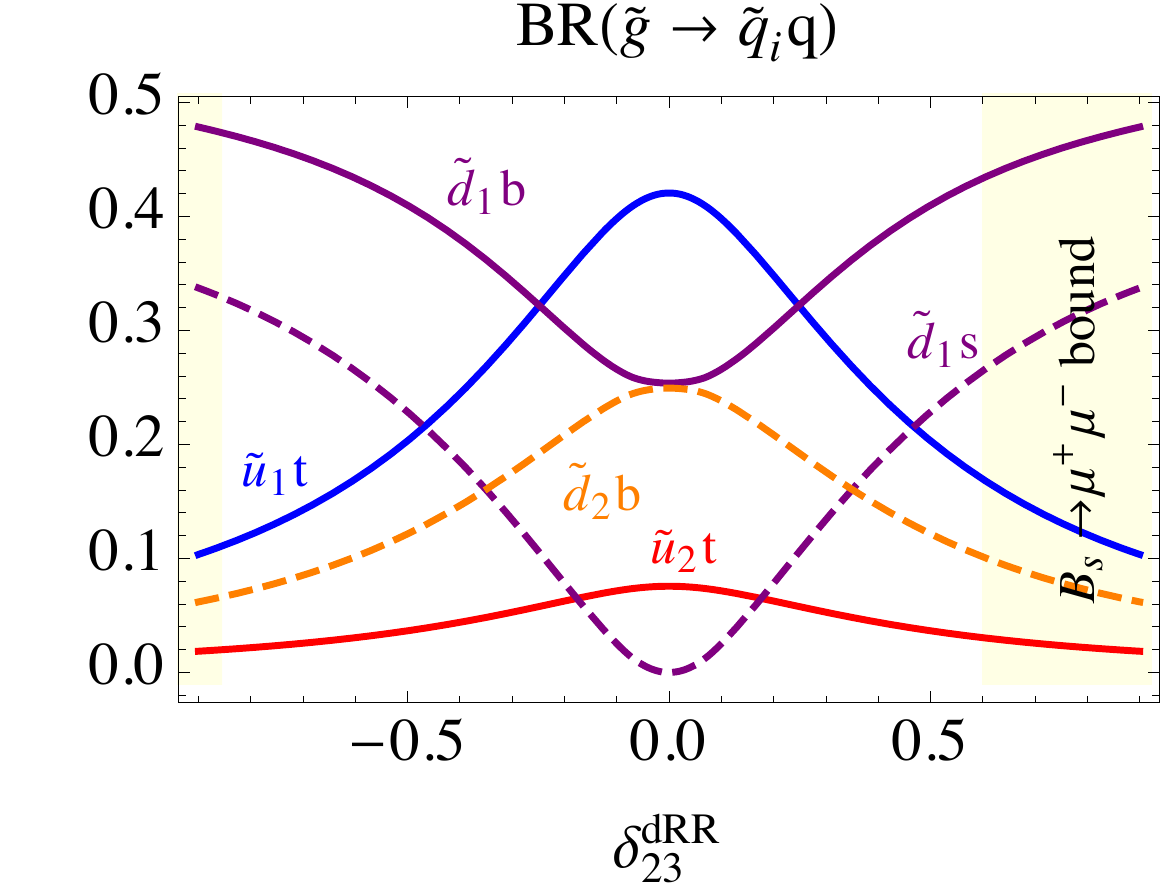}} \hspace*{-1cm}}}
  \label{fig2d}}
\caption{(a) Total two-body decay width $\Gamma(\sg \to \sq q)$ at
tree-level and full one-loop corrected (which coincides with the
SQCD one-loop corrected one) as functions of the QFV parameter $\ddrr$;
(b) $\Delta \Gamma(\sg \to \sq q)$ being the SQCD one-loop and the
full one-loop corrections to $\Gamma(\sg \to \sq q)$ relative to
the tree-level width;
(c) Partial decay widths and (d) branching ratios of the kinematically
allowed individual two-body channels at full one-loop level
as functions of $\durr$.
All the other parameters are fixed as in Table~\ref{basicparam}, except
$\durl = \dulr =\durr = 0$.}
\label{fig2}
\end{figure*}
In Fig.~\ref{fig2} we show dependences on the QFV parameter $\ddrr$. In~\ref{fig2a} 
the tree-level, the SQCD and total full one-loop widths 
and in~\ref{fig2b} the relative contribution of the one-loop SQCD and the full one-loop part in terms of the tree-level result are shown. 
The partial decay widths as well as the branching ratios of the kinematically allowed two-body channels are shown in~\ref{fig2c} and~\ref{fig2d}, respectively.
A comparison of Fig.~\ref{fig2} with Fig.~\ref{fig1} demonstrates the equal importance of QFV mixing in both $\su$ and $\sd$ sector. But in the 
$\sd$ sector all plots are more symmetric around $\ddrr = 0$ compared to these in $\su$ sector around  $\durr = 0$. This stems from the fact that 
in the $\su$ mass matrix $T_{U33} = 2500$~GeV but in the $\sd$ mass matrix $T_{D33} = 0$~GeV is taken and $m_b\, \mu \tan\beta$ is relatively small,
see eq.~(\ref{RLblocks}).
The SQCD corrected width in~\ref{fig2a} seem to coincide with the full one-loop corrected width, which we see in detail in~\ref{fig2b}. 
There the SQCD correction is about -7.5\% and varies only within 1\% around this value. The EW part varies between -0.5\% to -1.5\%.
In the Figs.~\ref{fig2c} and \ref{fig2d} we see that for large absolute values of the $\sd$~right-right mixing parameter $\ddrr$ the $\sd_1$ decay modes become 
much more important than the $\su$ ones since the $\sd_1$ mass becomes smaller due to the mixing effect.
As $\sd_{1,2}$ are mainly bottom squarks, the EW corrections to
the $\sd_{1,2} b$ modes are small,
mainly controlled by the rather small bottom-quark Yukawa coupling
$Y_b(Q = 3~{\rm TeV})$ for $\tan\b = 15$. On the other hand,
as $\su_{1,2}$ are mainly top squarks, the EW corrections to
the $\su_{1,2} t$ modes are significant,
mainly controlled by the large top-quark Yukawa coupling $Y_t$.
This explains the smallness of the EW corrections in \ref{fig2a} and \ref{fig2b},
especially for large $|\ddrr|$.\\

Fig.~\ref{fig2D1} shows the relative contribution of the one-loop SQCD and the full one-loop part in terms of the tree-level result for the 
partial decay width~\ref{fig2D1a} and the branching ratio~\ref{fig2D1b} of the decay $\tilde g \to \tilde u_1 t$ as a function of $\ddrr$ in the phenomenologically allowed region. The interesting point is that the dependence of this channel on $\ddrr$ comes mainly from the gluino wave function correction term with $\tilde d$ in the loop. The SQCD correction varies between -8\% and - 9.5\% and the EW correction is about constant and is $\sim$~-3\% for the width~\ref{fig2D1a}. For the 
branching ratio~\ref{fig2D1b}, the effect is much smaller for the SQCD correction, between -0.5\%  and -1.5\%. The EW part is maximal -3\%.\\
\begin{figure*}[h!]
\centering
\subfigure[]{
   { \mbox{\hspace*{-1cm} \resizebox{7.5cm}{!}{\includegraphics{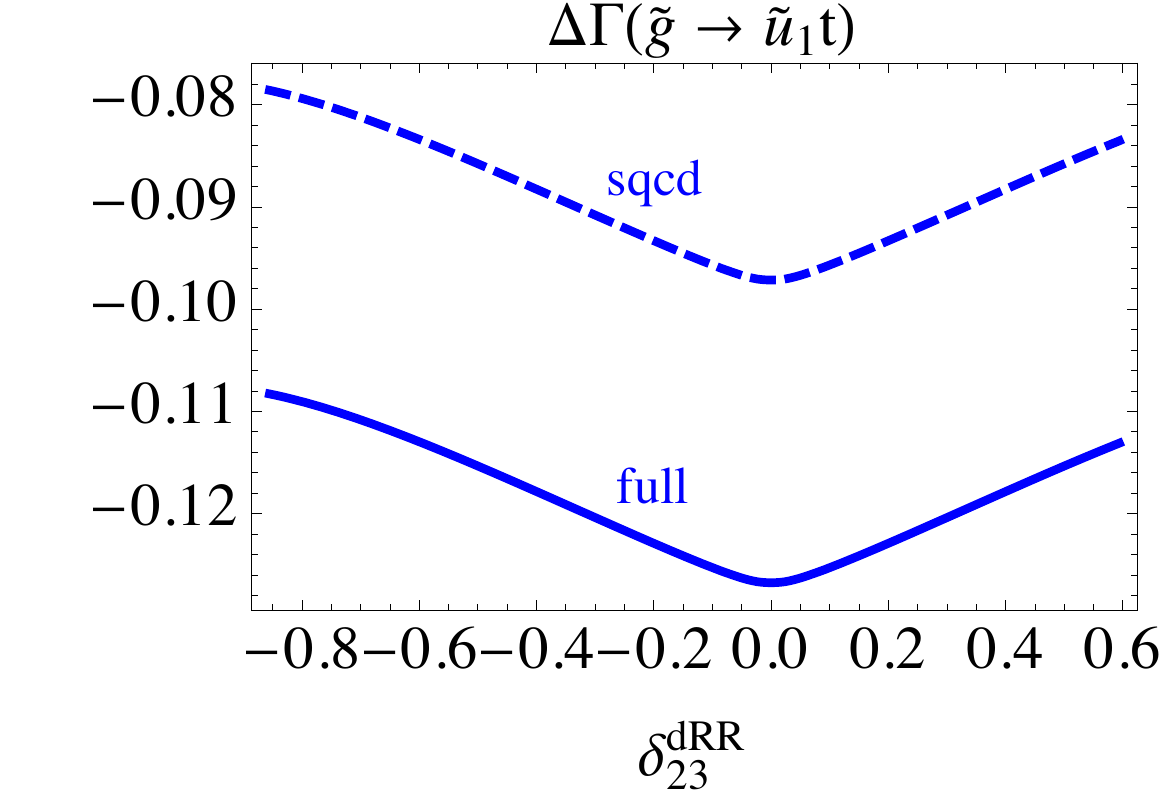}} \hspace*{-0.8cm}}}
   \label{fig2D1a}}
\subfigure[]{
   { \mbox{\hspace*{0.5cm} \resizebox{7.5cm}{!}{\includegraphics{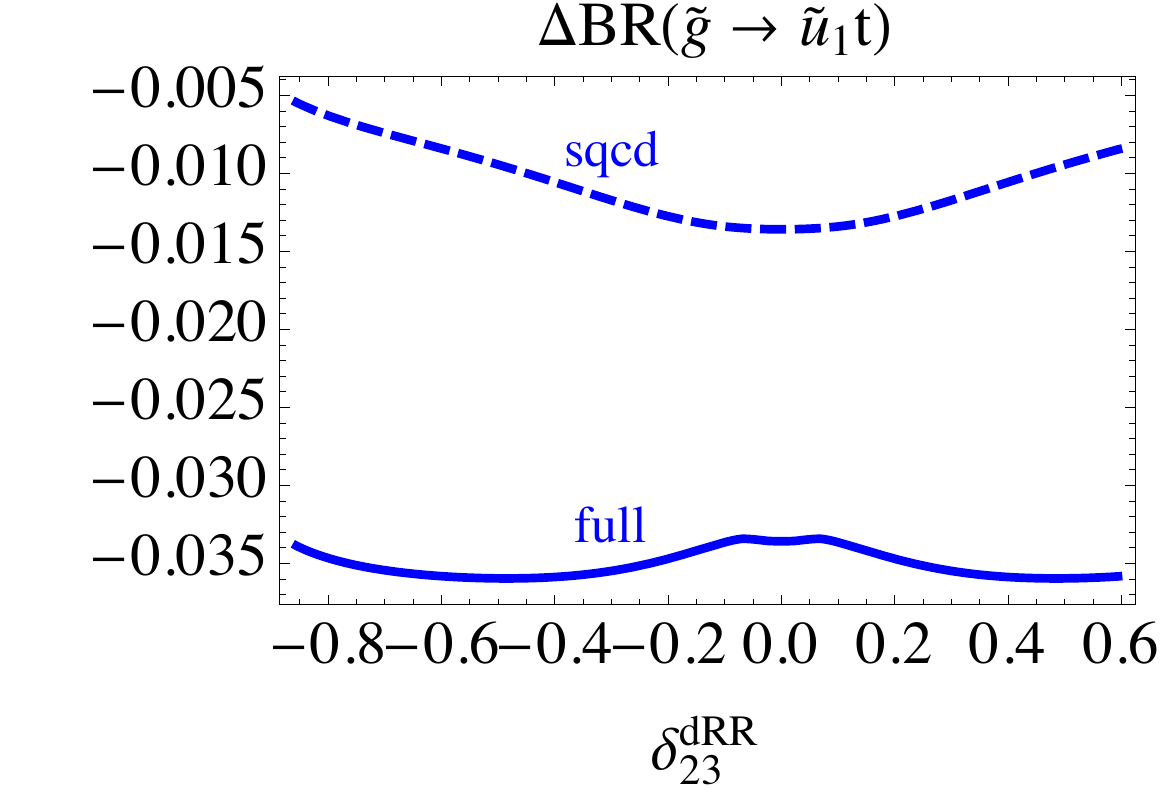}} \hspace*{-0.8cm}}}
   \label{fig2D1b}} 
\caption{$\Delta \Gamma$ and $\Delta$BR denote  the SQCD and the full one-loop contribution in terms of the tree-level result
for the decay $\sg \to \su_1 t$ as a function of $\ddrr$, (a) to the partial width, (b) to the branching ratio, respectively.
The parameters are fixed as in Fig.~\ref{fig2}.}
\label{fig2D1}
\end{figure*}
%
\begin{figure*}[h!]
\centering
\subfigure[]{
   { \mbox{\hspace*{-1cm} \resizebox{8cm}{!}{\includegraphics{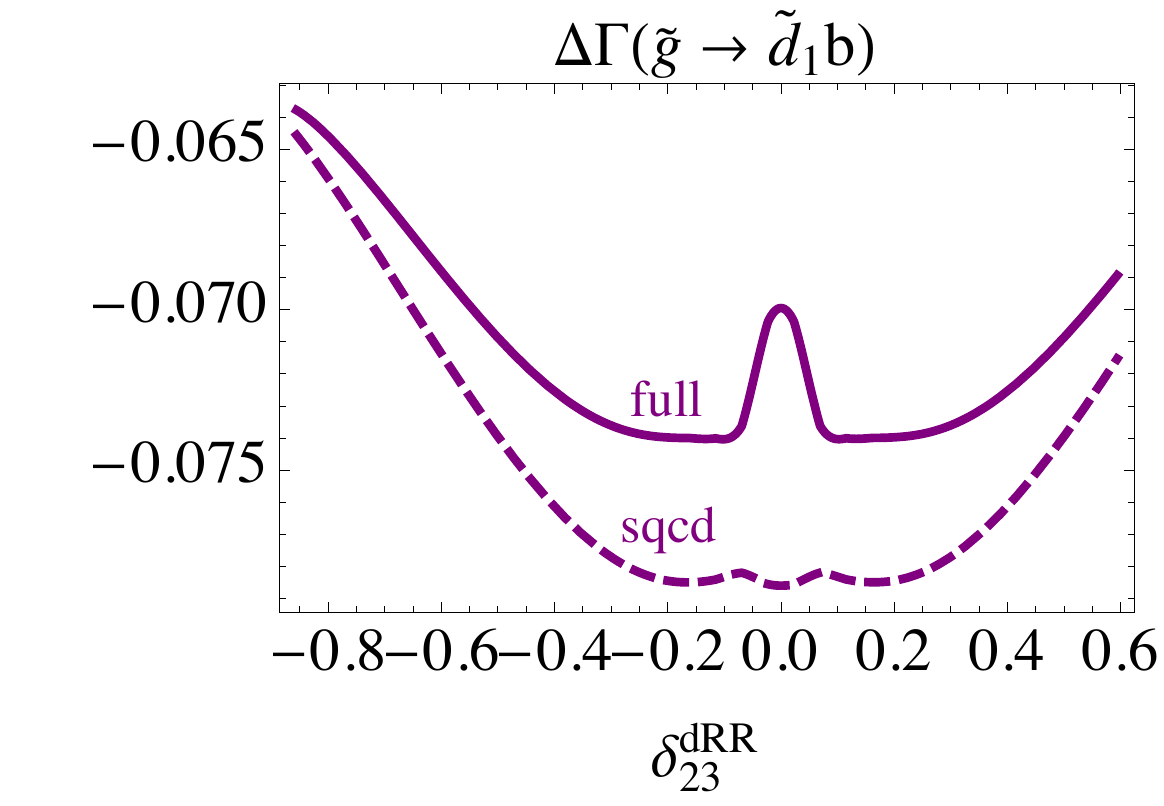}} \hspace*{-0.8cm}}}
   \label{fig2D2a}}
\subfigure[]{
   { \mbox{\hspace*{0.5cm} \resizebox{7.5cm}{!}{\includegraphics{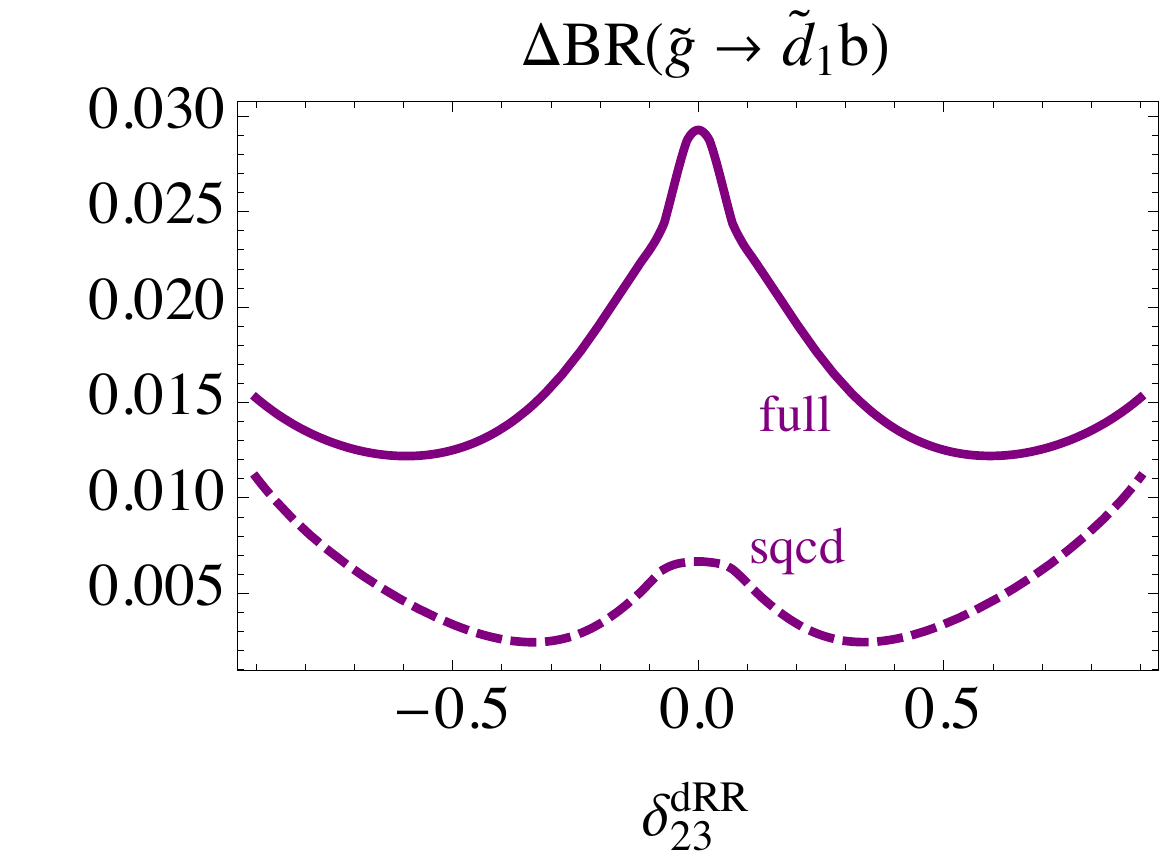}} \hspace*{-0.8cm}}}
   \label{fig2D2b}} 
\caption{$\Delta \Gamma$ and $\Delta$BR denote  the SQCD and the full one-loop contribution in terms of the tree-level result
for the decay $\sg \to \sd_1 b$ as a function of $\ddrr$, (a) to the partial width, (b) to the branching ratio, respectively.
The parameters are fixed as in Fig.~\ref{fig2}.}
\label{fig2D2}
\end{figure*}

Fig.~\ref{fig2D2} shows the relative contribution of the one-loop SQCD and the full one-loop part in terms of the tree-level result for the 
partial decay width~\ref{fig2D2a} and the branching ratio~\ref{fig2D2b} of the decay $\tilde g \to \tilde d_1 b$ as a function of $\ddrr$ in the phenomenologically allowed region. The SQCD correction varies between -6.5\% and - 8\% and the EW correction can become $\sim$~1\% for the width~\ref{fig2D2a}. 
For the branching ratio~\ref{fig2D2b}, the effects are again smaller, the SQCD correction is less than 1\% and the EW part maximal 3\%.
As in Fig.~\ref{fig1D} the wiggles stem from the complex structures of the QFV one-loop contributions.\\

In Fig.~\ref{fig3} a simultaneous dependence on the right-right mixing parameters of both $\su$ and $\sd$ sectors is shown. It is clearly seen that the total two-body decay width  $\Gamma(\sg \to \sq q)$ can vary up to 70 GeV in the allowed parameter region due to QFV.\\
 %
\begin{figure*}[ht!]
\centering
   { \mbox{\hspace*{-1cm} \resizebox{8.cm}{!}{\includegraphics{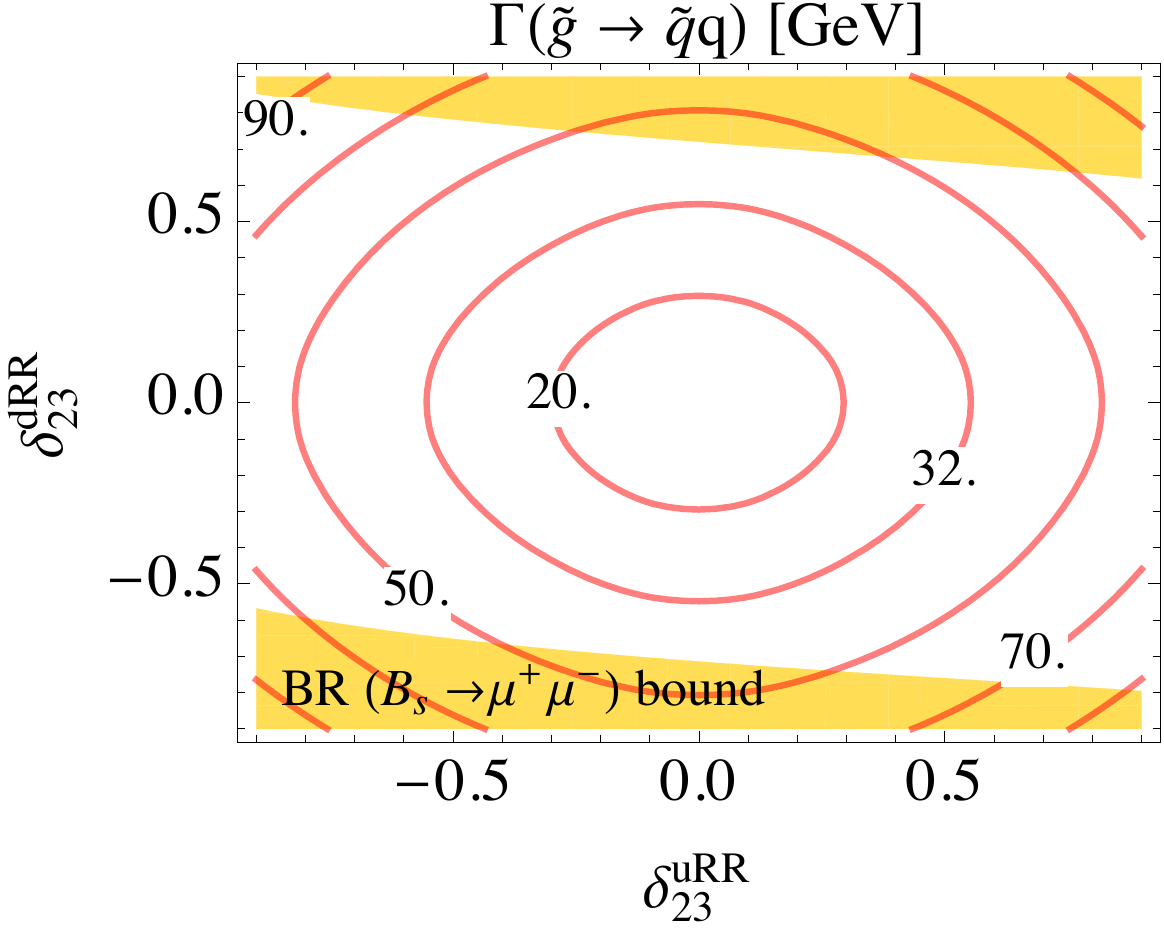}} \hspace*{-0.8cm}}}
\caption{Total two-body decay width $\Gamma(\sg \to \sq q)$ at full one-loop level as a function of the QFV parameters $\ddrr$ and $\durr$.
All the other parameters are given in Table~\ref{basicparam}, except $\durl = \dulr= 0.01$.}
\label{fig3}
\end{figure*}

In Fig.~\ref{fig3Da} the full one-loop part in terms of the tree-level result and in~\ref{fig3Db} the EW contribution relative to the SQCD contribution are shown
for the total two-body gluino decay width as a function of $\durr$ and $\ddrr$. We see in~\ref{fig3Da} a constant QFC one-loop contribution of $\sim$ -10\% 
and $\sim$~3\% variation due to QFV. 
The EW part can become up to $\sim$~35\% of the SQCD one (\ref{fig3Db})
for large $|\durr|$ where the $\su_1 t$ mode becomes important,
since the $\su_1$ mass becomes smaller due to the $\su$-sector
right-right mixing effect. Furthermore, as $\su_1$ is mainly a top squark,
the EW corrections to the $\su_1 t$ mode are significant, mainly controlled by the large top-quark
Yukawa coupling $Y_t$.\\

\begin{figure*}[ht!]
\centering
\subfigure[]{
   { \mbox{\hspace*{-1cm} \resizebox{8.cm}{!}{\includegraphics{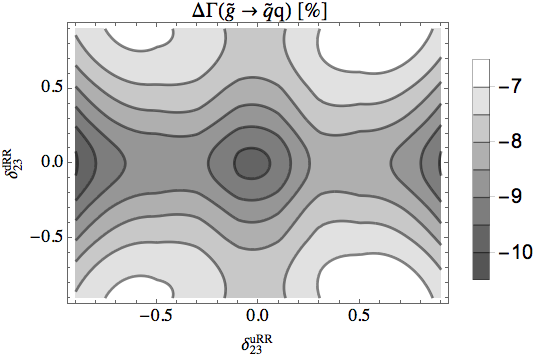}} \hspace*{0.5cm}}}
    \label{fig3Da}}
\subfigure[]{
   { \mbox{\hspace*{-1cm} \resizebox{8.cm}{!}{\includegraphics{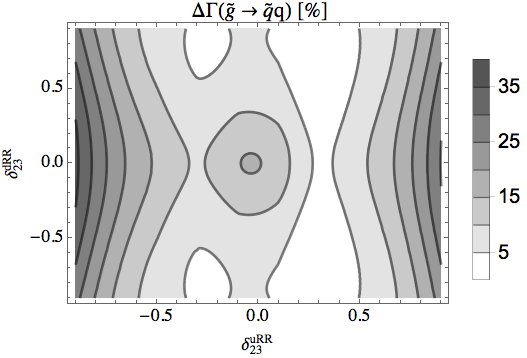}} \hspace*{-0.8cm}}}
    \label{fig3Db}}   
\caption{$\Delta \Gamma$ denotes in (a) the full one-loop contribution in terms of the total tree-level width,
in (b) the EW contribution relative to the SQCD contribution. Both plots are given as a   
function of the QFV parameters $\durr$ and $\ddrr$.
All the other parameters are given in Table~\ref{basicparam}, except $\durl = \dulr= 0.01$.}
\label{fig3D}
\end{figure*}
Fig.~\ref{figmgl} shows the dependence of the total two-body decay width $\Gamma(\sg \to \sq q)$ on the gluino mass in our reference scenario~\ref{figmgla} and in a quark-flavour conserving scenario, setting all QFV ($\d$) parameters of Table~\ref{basicparam} to zero~\ref{figmglb}. It is seen that in the QFV scenario~\ref{figmgla} $\Gamma(\sg \to \sq q)$ is somewhat enhanced. 
Because of the large $|\durr|$ ($|\ddrr|$) the mass difference between 
$\su_{1}$ and $\su_{2}$ ($\sd_{1}$ and $\sd_{2}$) is bigger. Consequently, $\su_1$ and $\sd_1$ are lighter and decays into these particles are already possible for smaller gluino masses.\\
\begin{figure*}[ht!]
\centering
\subfigure[]{
   { \mbox{\hspace*{-1cm} \resizebox{7cm}{!}{\includegraphics{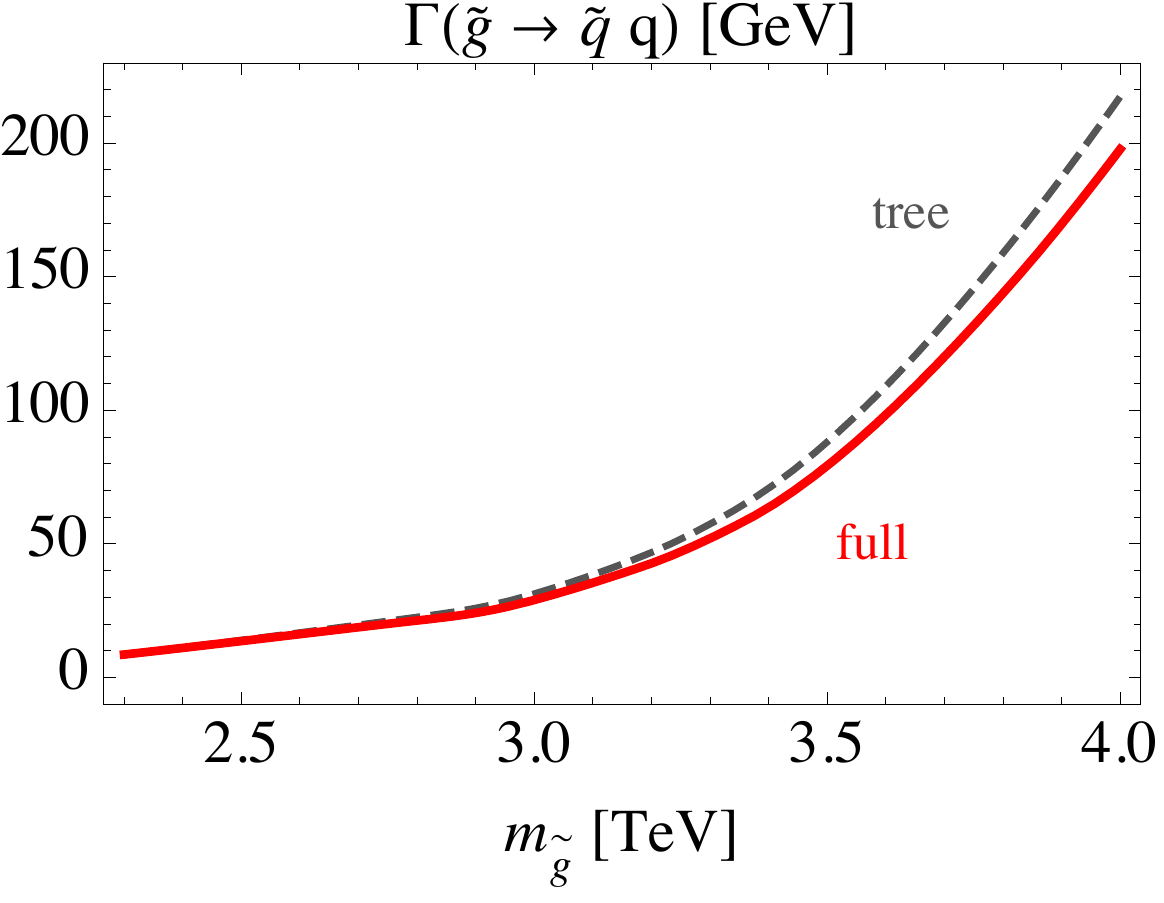}} \hspace*{-0.8cm}}}
   \label{figmgla}}
 \subfigure[]{
   { \mbox{\hspace*{+0.5cm} \resizebox{7.cm}{!}{\includegraphics{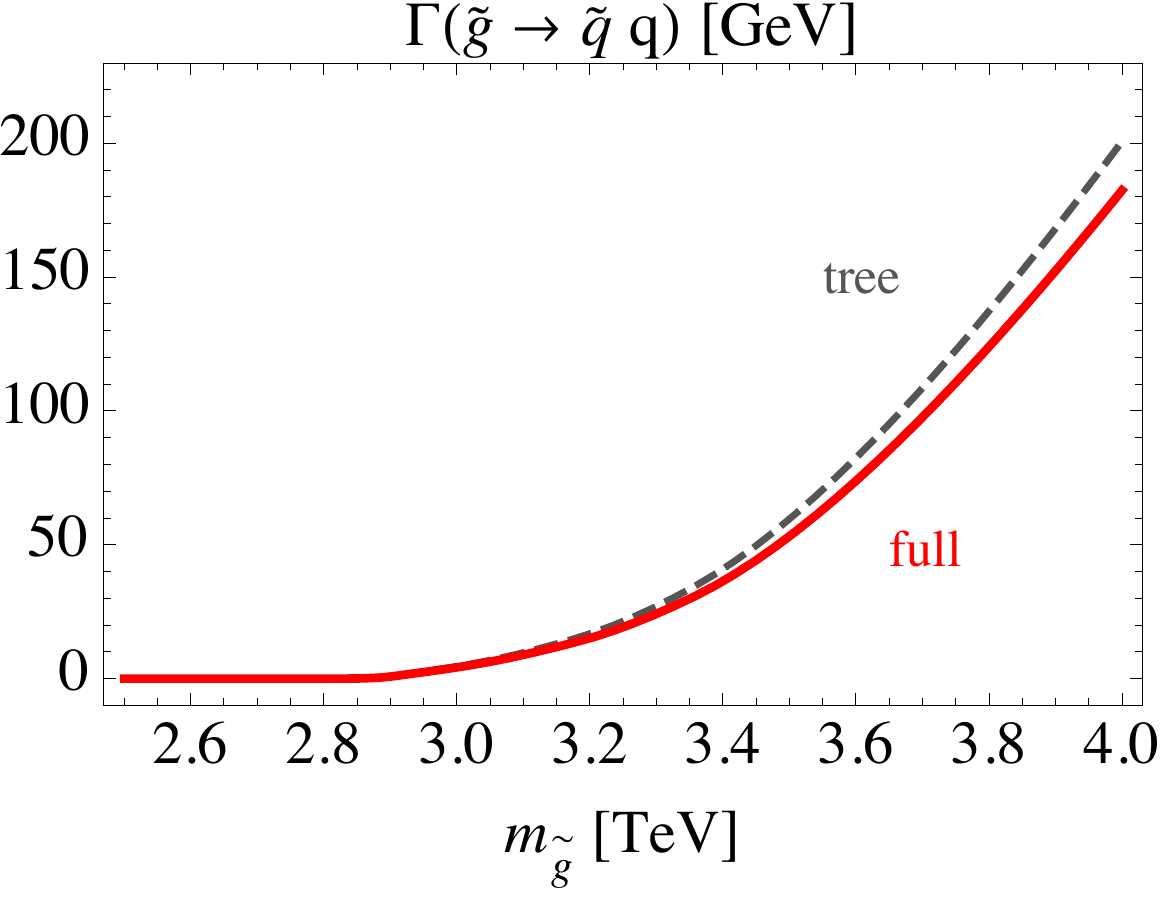}} \hspace*{-1cm}}}
  \label{figmglb}}\\
\caption{Dependence of the total two-body decay width $\Gamma(\sg \to \sq q)$  at tree-level (dashed) and full one-loop level (solid) on the gluino mass. (a)  QFV scenario with the parameters as given in Table~\ref{basicparam}; (b) QFC scenario with the parameters as given in Table~\ref{basicparam}, but with all QFV ($\d$) parameters set to zero.
}
\label{figmgl}
\end{figure*}

We have compared our numerical results in the flavour conserving limit with the results obtained in~\cite{Heinemeyer:2011ab}. 
For their reference scenario with $M_3 = 2000$~GeV assuming their input parameters to be $\overline {\rm DR}$ ones, 
we get a total width of 379~GeV. We agree with them within 2\%. For the relative size of the full one-loop correction we get -2\%
compared to their result of -2.5\%.

\section{Conclusions}
\label{sec:concl}

We have studied all two-body decays of the gluino at full one-loop level 
in the Minimal Supersymmetric Standard Model with quark-flavour violation 
in the squark sector. 
We have discussed a scenario 
where only the decays to $\tilde u_{1,2}$ and $\tilde d_{1,2}$ are 
kinematically open and $\tilde u_1$ is a mixture of $\tilde c_R$ 
and $\tilde t_R$ controlled by $\delta_{23}^{uRR}$, and $\tilde d_1$ 
is a mixture of $\tilde s_R$ and $\tilde b_R$ controlled by 
$\delta_{23}^{dRR}$. All other QFV parameters are small in order 
to fulfil the constraints from B-physics. 
The LHC constraints for the masses of the SUSY particles
are also satisfied, especially that one for $m_{h^0}$
and the vacuum stability conditions are fulfilled. 

The full one-loop corrections to the gluino decay widths are mostly 
negative. For the total decay width they are in the range of -10\%  
with a weak dependence on QFV parameters for both SQCD (including 
gluon loops) and electroweak (including also photon loops) 
corrections. For the decay width into $\su_1$ we can have a  total correction
up to -18\%, with the EW part up to -8\%, 
strongly depending on the QFV parameters. For the corresponding 
branching ratio the effect is somehow washed out. 
For the decay into $\sd_1$ we have maximal corrections of -8\%. 
In general, it turns out that the EW corrections can be 
in the range of up to 35\% of the SQCD corrections due to the large top-quark Yukawa coupling. 
The full one-loop corrections to the total width are of the order of
about -10\% in the gluino mass range of 2.3 - 4.0 TeV.

%
\section*{Acknowledgments}

We thank Sebastian Frank for providing the program FVSFOLD.
This work is supported by the "Fonds zur F\"orderung der
wissenschaftlichen Forschung (FWF)" of Austria, project No. P26338-N27.

\begin{appendix}

\section{Interaction Lagrangian}
\label{sec:lag}
The interaction of gluino, squark and quark is given by
\bea
{\cal L}_{\sg \sq_i q_g}&=& -\sqrt{2} g_s T_{rs}^{\a}\bigg[\bar{\sg}^{\a}(U^{\sq}_{i,g}e^{-i\frac{\phi_3}{2}}P_L-U^{\sq}_{i,g+3}e^{i\frac{\phi_3}{2}}P_R) q_g^s \sq_i^{*, r} \nn
&& \hspace*{2.2cm} +\bar{q}_g^r
(U^{\sq *}_{i,g} e^{i\frac{\phi_3}{2}}P_R-U^{\sq *}_{i,g+3}e^{-i\frac{\phi_3}{2}} P_L) \sg^{\a} \sq_i^s\bigg]\,,
\eea
where $T^\a$ are the SU(3) colour group generators, g is the generation index ($g=u,c,t$ for up-type quarks and $g=d,s,b$ for down-type quarks), and summation over $r,s=1,2,3$ and over $\a=1,...,8$ is understood. In our case the parameter $M_3=\msg e^{i \phi_3}$ is taken to be real, {\it i.e.} $\phi_3=0$.

\section{Theoretical and experimental constraints}
\label{sec:constr}

Here we summarize the experimental and theoretical constraints taken into 
account in the present paper. 
The constraints on the MSSM parameters from the B-physics experiments and 
from the Higgs boson measurement at LHC are shown in Table~\ref{TabConstraints}.
The constraints from the decays $B\to D^{(*)}\,\tau\,\nu$ are unclear due to 
large theoretical uncertainties \cite{Bartl:2014bka}. Therefore, we do not take 
these constraints into account in our paper.
In \cite{Dedes_t2ch} it is shown that the QFV decay $t \to c\,h^0$ in the 
current LHC runs cannot give any significant constraint on the 
$\tilde c - \tilde t$ mixing.

For the mass of the Higgs boson $h^0$, taking the combination of the ATLAS and 
CMS measurements  $m_{h^0} = 125.09 \pm 0.24~\gev$ \cite{Higgs_mass_ATLAS_CMS} and 
adding the theoretical uncertainty of $\sim \pm 3~\gev$ ~\cite{Higgs_mass_Heinemeyer}
linearly to the experimental uncertainty at 2 $\sigma$, 
we take $m_{h^0} = 125.09 \pm 3.48 ~\gev$.
%
\begin{table*}[t]
\footnotesize{
\caption{
Constraints on the MSSM parameters from the B-physics experiments
relevant mainly for the mixing between the second and the third generations of 
squarks and from the data on the $h^0$ mass. The fourth column shows constraints 
at $95 \%$ CL obtained by combining the experimental error quadratically
with the theoretical uncertainty, except for $m_{h^0}$.
}
\begin{center}
\begin{tabular}{|c|c|c|c|}
    \hline
    Observable & Exp.\ data & Theor.\ uncertainty & \ Constr.\ (95$\%$CL) \\
    \hline\hline
    &&&\\
    $\Delta M_{B_s}$ [ps$^{-1}$] & $17.757 \pm 0.021$ (68$\%$ CL)~\cite{DeltaMBs_HFAG2014} 
    & $\pm 3.3$ (95$\%$ CL)~\cite{DeltaMBs_Carena2006, Ball_2006} &
    $17.757 \pm 3.30$\\
    $10^4\times$B($b \to s \gamma)$ & $3.41 \pm 0.155$ (68$\%$ CL)~\cite{Trabelsi_EPS-HEP2015} 
    & $\pm 0.23$ (68$\%$ CL)~\cite{Misiak_2015} &  $3.41\pm 0.54$\\
    $10^6\times$B($b \to s~l^+ l^-$)& $1.60 ~ ^{+0.48}_{-0.45}$ (68$\%$ CL)~\cite{bsll_BABAR_2014}
    & $\pm 0.11$ (68$\%$ CL)~\cite{Huber_2008} & $1.60 ~ ^{+0.97}_{-0.91}$\\
    $(l=e~{\rm or}~\mu)$ &&&\\
    $10^9\times$B($B_s\to \mu^+\mu^-$) & $2.8~^{+0.7}_{-0.6}$ (68$\%$CL)~\cite{Bsmumu_LHCb_CMS}
    & $\pm0.23$  (68$\%$ CL)~\cite{Bsmumu_SM_Bobeth_2014} 
    & $2.80~^{+1.44}_{-1.26}$ \\
    $10^4\times$B($B^+ \to \tau^+ \nu $) & $1.14 \pm 0.27$ (68$\%$CL)
    ~\cite{Trabelsi_EPS-HEP2015, Hamer_EPS-HEP2015}
    &$\pm0.29$  (68$\%$ CL)~\cite{Btotaunu_LP2013} & $1.14 \pm 0.78$\\
    $ m_{h^0}$ [GeV] & $125.09 \pm 0.24~(68\%~ \rm{CL})$ \cite{Higgs_mass_ATLAS_CMS}
    & $\pm 3$~\cite{Higgs_mass_Heinemeyer} & $125.09 \pm 3.48$ \\
&&&\\
    \hline
\end{tabular}
\end{center}
\label{TabConstraints}}
\end{table*}
%

In addition to these constraints we also require our scenarios to be  
consistent with the following experimental constraints: 

(i) The LHC limits on the squark and gluino masses (at 95\% CL) 
~\cite{SUSY@ICHEP2016}:

In the context of simplified models, gluino masses $\msg \lesssim 1.9~{\rm TeV}$ are 
excluded at 95\% CL. The mass limit varies in the range 1400-1900~GeV depending 
on assumptions. First and second generation squark masses are excluded below 1400~GeV. 
Bottom squark masses are excluded below 1000~GeV. A typical top-squark mass limit is 
$\sim$ 900~GeV. 

(ii) The LHC limits on $m_{\ch_1}$ and $m_{\nt_2}$ from negative
searches for charginos and neutralinos mainly in leptonic final states
~\cite{SUSY@ICHEP2016}.

(iii) The constraint on ($ m_{A^0, H^+} , \tan \b $) from the MSSM Higgs boson
searches at LHC ~\cite{ICHEP2016_ATLAS, Charged_Higgs@ATLAS}.

(iv) The experimental limit on SUSY contributions on the electroweak
$\rho$ parameter ~\cite{Altarelli:1997et}: $\Delta \rho~ (\rm SUSY) < 0.0012.$

Furthermore, we impose the following theoretical constraints from the vacuum 
stability conditions for the trilinear coupling matrices~\cite{Casas}: 
\begin{eqnarray}
|T_{U\alpha\alpha}|^2 &<&
3~Y^2_{U\alpha}~(M^2_{Q \alpha\alpha}+M^2_{U\alpha\alpha}+m^2_2)~,
\label{eq:CCBfcU}\\[2mm]
|T_{D\alpha\alpha}|^2 &<&
3~Y^2_{D\alpha}~(M^2_{Q\alpha\alpha}+M^2_{D\alpha\alpha}+m^2_1)~,
\label{eq:CCBfcD}\\[2mm]
|T_{U\alpha\beta}|^2 &<&
Y^2_{U\gamma}~(M^2_{Q \beta\beta}+M^2_{U\alpha\alpha}+m^2_2)~,
\label{eq:CCBfvU}\\[2mm]
|T_{D\alpha\beta}|^2 &<&
Y^2_{D\gamma}~(M^2_{Q \beta\beta}+M^2_{D\alpha\alpha}+m^2_1)~,
\label{eq:CCBfvD}
\end{eqnarray}
where
$\a,\b=1,2,3,~\a\neq\b;~\gamma={\rm Max}(\a,\b)$ and
$m^2_1=(m^2_{H^+}+m^2_Z\sin^2\theta_W)\sin^2\b-\frac{1}{2}m_Z^2$,
$m^2_2=(m^2_{H^+}+$  
$m^2_Z\sin^2\theta_W)$ $\cos^2\beta-\frac{1}{2}m_Z^2$.
The Yukawa couplings of the up-type and down-type quarks are
$Y_{U\alpha}=\sqrt{2}m_{u_\alpha}/v_2=\frac{g}{\sqrt{2}}\frac{m_{u_\alpha}}{m_W
\sin\beta}$
$(u_\a=u,c,t)$ and
$Y_{D\alpha}=\sqrt{2}m_{d_\alpha}/v_1=\frac{g}{\sqrt{2}}\frac{m_{d_\alpha}}{m_W
\cos\beta}$
$(d_\a=d,s,b)$,
with $m_{u_\a}$ and $m_{d_\a}$ being the running quark masses at the weak
scale and $g$ being the SU(2) gauge coupling. All soft SUSY-breaking parameters 
are given at $\rm Q=3$~TeV. As SM parameters we take $m_Z=91.2~\gev$ and
the on-shell top-quark mass $m_t=173.3~\gev$ \cite{top_mass@ICHEP2014}.

\section{Hard photon/gluon radiation}
\label{appendix_hardradiation}

\begin{figure}[h!]
\caption{The combination of three Feynman graphs for the $1 \to 3$ bremsstrahlung process emitting a 
photon or a gluon from a fermion to scalar-fermion structure.
\label{feyn_brems_generic}}
\begin{center}
\mbox{\resizebox{12cm}{!}{\includegraphics{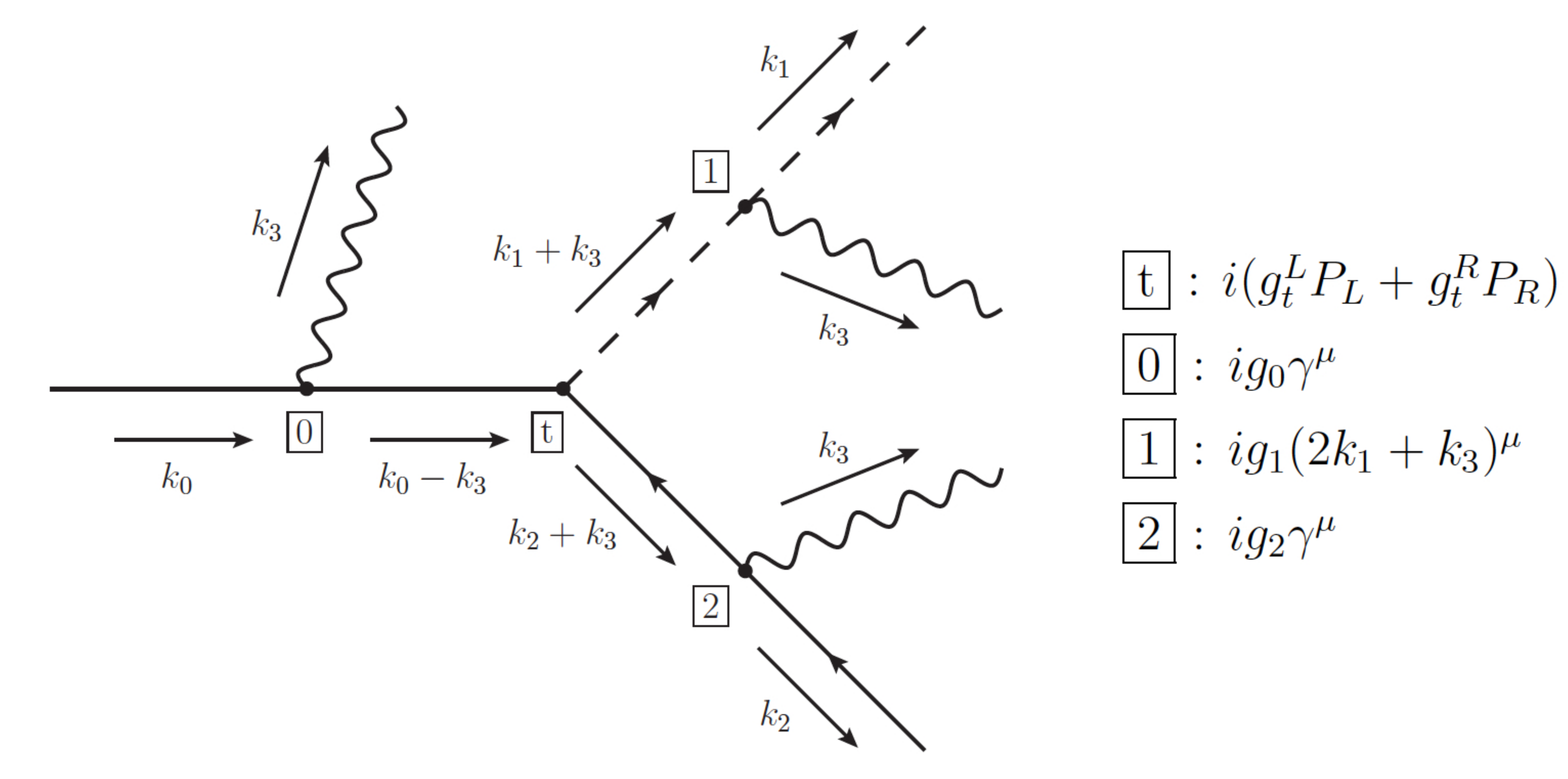}}}
\end{center}
\end{figure}

We start with the general formula of a $1 \to 3$ process with the hard radiation of a photon or a gluon,
\begin{equation}
\Gamma^{\rm hard} = {1 \over 2^6 m_0 \pi^3} \int {{\rm d}^3k_1 \over 2 E_1} {{\rm d}^3k_2 \over 2 E_2}  {{\rm d}^3k_3 \over 2 E_3} \delta^4(k_0 - k_1 - k_2 - k_3)
\overline{|{\cal M}^{\rm hard} |^2}\, .
\label{gamma_brems0}
\end{equation}
The bar means we take the average of incoming spins and colours and sum over the outgoing spins and colours. Based on the 
diagram Fig.~\ref{feyn_brems_generic} and using the definition of the bremsstrahlung's integrals from \cite{Denner93} we can write eq.~(\ref{gamma_brems0}) as
\begin{equation}
\Gamma^{\rm hard} = {col \over 2^6 m_0 \pi^3} X_{\rm FSF}\, ,
\label{gamma_brems}
\end{equation}
where $col$ denotes the colour average of the incoming particle and the fermion to scalar-fermion structure factor
\bea
X_{\rm FSF} &=& g_0^2 \bigg [ \left( -2 \a m_0^2 -2 \b m_2 m_0 \right) I_0 - \a I_0^2 \nn
&+& \left( -2 \a \left( m_0^2 -m_1^2 +m_2^2\right) m_0^2 - 4 \b m_2 m_0^3 \right) I_{00}\bigg ] \nn
&+& g_0 g_1 \bigg [ -\a I +\left( 2 \a(m_1^2-m_2^2) - 2 \b m_0 m_2 \right) I_0\nn 
&+& \left( \a (-m_0^2-m_1^2-m_2^2) - 2 \b m_0 m_2 \right) I_1\nn
&+& \left( 2 \a \left( (m_1^2-m_2^2)^2-m_0^4 \right) -4 \b m_0 m_2  (m_0^2+m_1^2-m_2^2)\right)  I_{10} \bigg]\nn
&+& g_1^2 \bigg[ \a I +\left( \a (-m_0^2+3 m_1^2-m_2^2) - 2 \b m_0 m_2 \right) I_1\nn 
&+& \left( -2\a (m_0^2-m_1^2+m_2^2) m_1^2 -4 \b m_0 m_2 m_1^2\right) I_{11}\bigg]\nn
&+& g_0 g_2 \bigg [ -2 \a I +\left( 2 \a(m_1^2-m_2^2) - 2 \b m_0 m_2 \right) I_0\nn
&+& \left( -2\a (m_0^2-m_2^2) - 2 \b m_0 m_2 \right) I_2\nn
&+& \left(- 2 \a (m_0^2-m_1^2+m_2^2)^2  -4 \b m_0 m_2  (m_0^2-m_1^2+m_2^2)\right)  I_{20} \bigg]\nn
&+& g_1 g_2 \bigg [ \a I +\left( \a(m_0^2+m_1^2+m_2^2)+2 \b m_0 m_2 \right) I_1\nn
&+& \left( 2\a (m_0^2-m_1^2)+ 2 \b m_0 m_2 \right) I_2\nn
&+& \left(  \a \left( 2 m_2^4-2(m_0^2-m_1^2)^2 \right)+4 \b m_0 m_2  (-m_0^2+m_1^2+m_2^2)\right)  I_{21} \bigg]\nn
&+& g_2^2 \bigg[ \a I + (-2 \a m_2^2 - 2 \b m_0 m_2 )I_2 +\a I_2^1\nn
&+& \left( -2\a (m_0^2-m_1^2+m_2^2) m_2^2 -4 \b m_0 m_2^3\right) I_{22}\bigg]\, ,
\label{hardrad}
\eea
where $\a = 2 g_s^2 (|U^{\sq}_{i,g}|^2+|U^{\sq}_{i,g+3}|^2)$ and $\b = - 4 g_s^2 {\rm Re}\big(U^{\sq *}_{i,g}\, U^{\sq}_{i,g+3} \big)$. 
Note, that the spin average for the incoming fermion of 1/2 is already included.
For the gluino decays $col$ is 1/8.

\noindent
The explicit result for photon radiation is
\begin{equation}
\Gamma(\sg \to \su_i u_g \gamma) = {1 \over 512 \pi^3 m_{\sg}}\, 4 \,X_{\rm FSF}\, ,
\end{equation}
taking in $X_{\rm FSF}$, eq.~(\ref{hardrad}), $g_0 = 0, g_1 = -e Q_1, g_2 = -e Q_2$, $e$ denotes the positron charge and $Q_{1,2}$ the charge of the particle 
on leg 1 or 2 in units of $e$, respectively. The addional factor 4 stems from the colour summation, which is universal, ${\rm Tr}(T^a T^a) = 3 C_F = 4$.\\

\noindent
The result for gluon radiation reads
\begin{equation}
\Gamma(\sg \to \su_i u_g g) = {1 \over 512 \pi^3 m_{\sg}} X_{\rm FSF}\, .
\end{equation}
In this case the colour summation results in the $3 \times 3$ matrix $C$. We 
take in $X_{\rm FSF}$, eq.~(\ref{hardrad}), $g_i g_j \to g_{si}g_{sj}C_{ij}$, where $g_{si}= g_s Q_{si}$, $g_s =\sqrt{4 \pi \a_s}$ is 
the strong coupling constant and $Q_{si}=\pm 1$ is the colour charge factor for particles carrying colour/anti-colour, respectively. 
The matrix $C$ describes the colour traces of the SU(3)$_C$ generators and is given by
\bea
C = \left( 
\begin{tabular}{c c c}
  \vspace*{0.3cm}
  12 & 6 & -6 \\
  \vspace*{0.3cm}
  6 & 16/3 & -2/3 \\
  -6 & -2/3 & 16/3
\end{tabular}
\right)\, ,
\eea
e.g. $C_{00} = 12$.


\end{appendix}

%
%


\end{document}